# Helicity-Resolved Spatiotemporal Mapping of Chiral Plexcitons in Helicoids

*Jeong Hyun Han[1†], Sankaran Ramesh[2†], Jaeyeon Jo[1,3], Pavel Chabera[2,4], Ryeong Myeong Kim[1,5], Sung Hoon Cho[1], In Han Ha[1], Amitav Sahu[2], Yoonsang Tak[1], Jiawei Lv[1], Miyoung Kim[1,3], Ki Tae Nam[1*], Tõnu Pullerits[2*]*

[1] Department of Materials Science and Engineering, Seoul National University, Seoul 08826, Republic of Korea

[2] Division of Chemical Physics and NanoLund, Lund University, 22100 Lund, Sweden

[3] Research Institute of Advanced Materials, Seoul National University, Seoul 08826, Republic of Korea

[4] Department of Physics, College of Science, United Arab Emirates University, P.O. Box 15551, Al Ain, United Arab Emirates

[5] Emerging Material Metrology Group, Division of Chemical and Material Metrology, Korea Research Institute of Standards and Science (KRISS), Daejeon 34113, Republic of Korea

**† These authors contributed equally to this work.**

*** To whom correspondence should be addressed:**

Ki Tae Nam, Ph.D.

Professor

Department of Materials Science and Engineering

Seoul National University

Seoul 08826, Republic of Korea

Tel: 82 2 880 7094

E-mail: nkitae@snu.ac.kr

Tönu Pullerits, Ph.D.

Professor

Division of Chemical Physics and NanoLund

Lund University

22100 Lund, Sweden

Tel: 46 46 222 81 31

E-mail: tonu.pullerits@chemphys.lu.se

## Abstract

Plasmon-exciton hybrids, or plexcitons, offer deeply subwavelength light-matter interactions with versatile pathways for energy redistribution. Incorporating chirality into such systems is particularly compelling, enabling spin-sensitive optical functionality that can operate on ultrafast timescales and within ultracompact volumes. Despite recent progress in chiral plexcitonic systems, how structural chirality and plasmon-exciton coupling determine chiroptical spectra and ultrafast energy flow remains elusive. Here we realize chiral plexcitons by functionalizing intrinsically chiral gold helicoid nanoparticles with molecular J-aggregates. Within a non-Hermitian framework, we trace the microscopic origin of the helicoid chiroptical response and its coupling to the excitonic transition, revealing how the helicity of light selectively addresses distinct hybrid responses. At the spatiotemporal extreme, we find that the gap-localized response not only enhances polarization-sensitive contrast but also strengthens the local hybrid interaction, leading to accelerated ultrafast relaxation. Together, these space-, time-, and polarization-resolved measurements provide a physically grounded and experimentally benchmarked picture of chiral plexcitonic coupling, identifying chirality as a practical control parameter for selectively steering nanoscale energy pathways and dynamics.

## Introduction

Engineering light-matter interaction at the nanoscale is central to quantum information processing, ultrasensitive sensing, and integrated photonic technologies[1,2]. At these dimensions, optical modes and material excitations no longer act independently; instead, the internal degrees of freedom of light, such as phase and polarization, play a decisive role in shaping the optical response and opening interaction pathways absent in bulk systems[3-5]. In this context, the spin degree of freedom of light, manipulated through its circular polarization, has emerged as a powerful handle for accessing chiral and spin-dependent phenomena[6-8]. Helicity acts as an active control parameter by changing the handedness of the optical excitation, thereby selectively coupling to interaction channels distinguished by chirality or spin, even within the same spectral band. Such advantages have motivated extensive efforts toward helicity-encoded functionality in a range of chiral material platforms, including chiral perovskite[9,10] and valley materials such as transition metal dichalcogenides[11,12]. While these systems enable helicity-selective access to internal electronic or excitonic states, their limited interaction strengths and spatial confinement are often insufficient to directly manipulate energy flow at the nanoscale[7,13].

Plexcitonic systems, formed by coupling excitonic transitions to plasmonic resonances, provide an effective route to overcome this limitation by combining strong near-field confinement with excitonic degrees of freedom[14-16]. Deep-subwavelength plasmonic mode volumes can greatly enhance light-matter interaction and imprint polarization into near-fields, motivating applications such as engineered energy-flow channels[17,18], polarization-selective addressing[19,20], and tailored emissive responses[21,22]. In the strong-coupling limit, coherent plasmon-exciton exchange manifests as spectral Rabi splitting, signalling the formation of controllable hybrid light-matter states[23,24]. However, plasmonic resonators are intrinsically broadband and lossy, so their coupling with narrowband excitons cannot always be described in terms of coherent energy exchange alone. Instead, it may also involve interference-driven response, often Fano-like in character, governed by distinct pathways of energy redistribution[25]. This ambiguity becomes even more pronounced in chiral plexcitonic platforms, where chiroptical effects complicate the interpretation of coupling physics and far-field circular-differential observables, such as circular dichroism (CD), can obscure the broadband plasmonic signatures typically used to diagnose coupling[26]. Although a range of chiral plexcitonic systems has recently been realized experimentally[27-29], many rely on only weakly broken symmetries and therefore lack a well-defined modal decomposition, hindering a quantitative theoretical description of the hybridization. Consequently, the fundamental microscopic picture of chiral plasmon-exciton coupling remains elusive, underscoring the need for quantitative models and experimental validation beyond steady-state observables to resolve the spatial distribution of hybrid states and their helicity-dependent ultrafast dynamics.

Here, we investigate helicity-dependent plasmon-exciton coupling in a hybrid plexcitonic system formed by functionalizing structurally chiral gold helicoid nanoparticles[30] with J-aggregate dyes. Helicoids provide a robust and experimentally accessible platform for chiral plexcitonic coupling because their morphology belongs to the chiral cubic point-group 432[31,32], which combines strong helicity selectivity with well-defined nanoscale localization of the relevant near-fields[30,33]. As a distinctive advantage of helicoids as a testbed, first, their clearly defined chiral morphological motifs support the concurrent excitation of plasmonic electric- and magnetic dipole modes (ED and MD) whose interference yields a pronounced and physically traceable chiroptical response across the chiroptically active spectral range[34,35]. This provides a solid basis for constructing a mechanistic model of plexcitonic coupling incorporating chirality. Second, this microscopic ED and MD basis naturally yields large far-field circular dichroism ($g = -0.2$) together with strong near-field optical helicity. Helicoid-based systems have indeed been demonstrated in coupled configurations with chiral molecules and quantum emitters, where the intrinsic chiroptical characteristics of the helicoids are retained and translated into the coupled responses[36-40]. Together, these features make helicoids an ideal platform for linking helicity-resolved far-field spectra with the underlying spatiotemporal characteristics of plexcitonic coupling.

Building on this experimental platform, we develop a non-Hermitian Hamiltonian (NHH) framework[41,42] that couples the plasmonic ED and MD modes to the excitonic transition, thereby capturing modal coupling, dissipation, and interference within a unified description. The analysis supports an interference-dominated, Fano-like coupling picture rather than one governed by coherent Rabi exchange, with direct implications for how energy is redistributed across the system. In particular, the plexciton response becomes spatially segregated into nanoscale domains with distinct spectral character: an upper branch (UB) weighted toward the helicoid periphery and a lower branch (LB) localized dominantly within the chiral gaps. The helicity of the incident light selectively addresses these domains, allowing polarization to function as a probe of energy relaxation in specific regions of the particle. On this basis, we find that plexcitonic coupling opens an accelerated energy-dissipation pathway within the chiral gaps. We validate this picture by combining nanoscale spatial mapping by electron energy loss spectroscopy (EELS) at individual particle level and femtosecond transient absorption (TA) with polarized pulses, unveiling how helicoid's morphology governs the localized spatiotemporal evolution of the hybrid states. More broadly, this work establishes a polarization-, time-, and space-resolved framework for analysing chiral plexcitonic coupling and points toward helicity-encoded control of energy flow in nanophotonic systems.

## Results

### Rational design of chiral plexcitonic helicoid system

Figure 1 summarizes the design of our chiral plexcitonic system, in which gold helicoids (I) are conformally functionalized with an excitonic shell of molecular J-aggregates (II). The two components were chosen to provide control over the spectral, spatial, temporal, and polarization-dependent characteristics of the hybrid response. Gold helicoids act as highly confined chiral plasmonic resonators, whose twisted concave gaps break the orthogonality between ED and MD responses and thereby generate strong local circular contrast within the nanoscale coupling region. This makes their spectral response directly accessible by polarization-resolved spectroscopy. The excitonic component is provided by TDBC, a cyanine dye that forms J-aggregates with a strong excitonic resonance at 590 nm and a large collective transition dipole moment[43,44]. Because this resonance overlaps well with the helicoid plasmon band and TDBC can be conjugated uniformly to the metal surface[16,45], it is well suited for probing plasmon-exciton coupling across the helicoid's intricate nanoscale morphology.

The coupled system (III) leverages the pronounced morphological and plasmonic chirality of the gold helicoid core to enable asymmetric, spatiotemporally selective excitation and readout of plexcitonic states via light polarization. The findings are as follows. Plasmon-exciton coupling produces spectral splitting into UB and LB (i). The incident polarization differentially excites these branches and their underlying modes, enabling selective access to the spatially inhomogeneous local density of states distributed between the chiral gaps and the particle periphery (ii, iii). This spatial asymmetry, rooted in the broken morphology of the helicoid, is further reflected in distinct ultrafast dynamics that depend on both energy and polarization (iv). Thus polarization serves as an active knob for selective probing and enhancing specific nanoscale pathways, allowing the spatiotemporal dissymmetry of the coupled system to be directly resolved.

**Realization of the helicoid-TDBC plexcitonic system**

Following this design rationale, we engineered the coupling conditions between the gold helicoids and TDBC to realize the plexcitonic system. Within an overall cubic outline, the helicoids exhibit three-dimensional structural distortions composed of chiral motifs associated with fourfold, threefold, and twofold symmetries, while maintaining remarkable morphological consistency across individual particles despite their highly twisted geometry (Figure 2a)[46]. Beyond the localized chiroptical response originating from the individual chiral gap motif, the helicoid's point-group symmetry 432, the most isotropic among the 11 chiral point groups[47], collectively integrates these local contributions across the entire particle. This collective effect drives an exceptionally robust net chiroptical response, with a large far-field Kuhn's dissymmetry factor of $g = -0.2$ (Figure 2b). This highly isotropic nature also ensures that the spatial localization of the plasmonic near-field remains highly consistent across individual particles, allowing the spatial characteristics of the coupled states to be reliably preserved even in bulk aqueous ensembles. In addition, the substantial spectral overlap between the excitonic resonance of TDBC J-aggregates and the helicoid LSPR band identified in the extinction spectrum establishes highly favourable conditions for efficient coupling (Supplementary Note 1). Direct probing of the plasmonic field via EELS further reveals a strong localized plasmonic mode (at 2.3 eV) that overlaps with the J-band, confirming its role in driving the ensuing plexcitonic coupling (Figure 2c)[48].

To induce and clearly resolve plexcitonic coupling in the helicoids, we employed the direct electrostatic assembly of negatively charged TDBC molecules onto positively charged CTAB-capped gold helicoids in aqueous colloidal solution[16,45]. At a fixed TDBC concentration, the resulting spectral features vary systematically with CTAB concentration, directly supporting this assembly mechanism (Figure 2d, Figures S3 and S4 in Supplementary Note 2)[49]. At low CTAB concentrations (< 100 μM), incomplete surface coverage hinders effective self-assembly, yielding an uncoupled TDBC monomer band at 518 nm. At intermediate concentrations (100 μM to 1 mM), $CTA^+$-assisted formation of floating J-aggregates becomes increasingly prominent, producing excess excitonic signals. Above the critical micelle concentration (> 1 mM), abundant CTAB micelles disrupt J-aggregation by trapping TDBC in its monomeric state, leading to the reappearance of the 518 nm band. Guided by the dissymmetry factor $g$, which selectively reports on plexcitonic coupling while excluding achiral TDBC contributions, we identified 100 μM CTAB as the optimal condition, where uncoupled species that would otherwise complicate spectroscopic analysis are minimized. Transmission electron microscopy energy-dispersive X-ray spectroscopy (TEM-EDX) elemental mapping further confirms successful conjugation under these optimized conditions (Figure S5 in Supplementary Note 2). This purification-free, direct one-pot approach provides a highly reproducible route to the plexcitonic helicoid system.

Taking advantage of the system's intrinsic chirality, incident helicity selectively addresses different hybrid responses of the coupled helicoid-TDBC system, as reflected in the distinct splitting observed in the $g$-factor and extinction spectra (Figure 2e). Under right circularly polarized (RCP) excitation, the spectral response exhibits a comparatively symmetric splitting profile that could, at first glance, be associated with the conventional picture of Rabi splitting. By contrast, under linear polarization (LP) or left circular polarization (LCP), the split response becomes increasingly asymmetric, indicating a change in the relative modal contributions and suggesting that interference plays an important role in shaping the optical lineshape. These helicity-dependent spectral variations therefore call for a more rigorous analysis of the coupling regime, which we present in the following section.

At the single-particle level, EELS provides independent evidence for plasmon-exciton coupling[50,51]. The clear loss splitting observed around 2.3 eV (Figure 2f) shows that the spectral splitting seen in optical measurements is not merely a far-field artifact. Moreover, the split EEL spectra vary systematically from the chiral gap to the particle periphery, indicating that the complex helicoid morphology gives rise to spatially differentiated hybrid response within a single particle, in addition to its polarization dependence (Figure S6 in Supplementary Note 3).

**Spectral analysis of the helicoid-TDBC plexcitonic response using a non-Hermitian model**

To characterize the coupling regime and clarify how incident polarization modulates the optical response, we investigated plasmon-exciton coupling from both macro- and microscopic perspectives. At the macroscopic level, the far-field spectroscopic signatures of plexcitonic interactions are particularly informative because they reflect the collective influence of near-field coupling within the nanoparticle. In helicoids, the chiroptical response has been shown to originate from the interference of two dipolar modes, ED and MD, with the latter arising from ED-associated current loops shaped by the particle's curved chiral morphology (Figure 3a,i; Supplementary Note 4). Using these modes as the basis of our description, we analyze how helicity selectively addresses the hybrid response and how the individual modal contributions evolve upon coupling to the excitonic transition. For this purpose, we employ a non-Hermitian Hamiltonian (NHH) model, which provides a unified framework for dissecting the experimental spectra and interpreting the resulting plexcitonic response[41,42].

We first introduce a 2×2 NHH for the plasmonic helicoids, written in the reduced basis of ED and MD contributions, to capture the dominant modal mixing underlying the system's chiroptical response:

$$\begin{pmatrix} E_E - i\gamma_E & V_{EM}e^{i\Phi_{EM}} \\ V_{EM}e^{-i\Phi_{EM}} & E_M - i\gamma_M \end{pmatrix} \tag{1}$$

where $E_E, E_M$ are the resonance energies of the $|ED\rangle$ and $|MD\rangle$ basis modes, while $\gamma_E, \gamma_M$ represent their linewidths associated with plasmon damping. The off-diagonal terms $V_{EM}e^{\pm i\Phi_{EM}}$ describe the effective mixing between these dipolar contributions, where $V_{EM}$ quantifies the interaction strength and $\Phi_{EM}$ represents the relative phase associated with the chiral geometry and near-field current distribution of the helicoid (Supplementary Note 5). Within this reduced description, the chiroptical response arises from the interplay of ED and MD contributions and their relative phase, providing an effective microscopic basis for the observed circular dichroism. In particular, the far-field CD can be expressed in terms of $V_{EM}$ and $e^{i\Phi_{EM}}$, in a form consistent with the conventional Rosenfeld description of optical activity (Supplementary Note 6)[52]. Upon diagonalization of the NHH, we obtain mixed transition dipole moments $(\mu_1, m_1)$ and $(\mu_2, m_2)$ for the two resulting modes from the basis moments $\mu_{ED}$ and $m_{MD}$. These are then used to calculate the linear optical attenuation under right- and left-circularly polarized illumination. By analytically decomposing the measured extinction spectra into Lorentzian components, we resolve the underlying modal contributions responsible for the helicity-dependent peak shifts and linewidth variations (Figure 3b, Figure S10 and Table S1 in Supplementary Note 5).

As a natural extension of this reduced description, we introduce a 3×3 NHH for the plexcitonic system by adding the excitonic transition and its effective interaction with the plasmonic ED and MD contributions:

$$\begin{pmatrix} E_E - i\gamma_E & V_{EM}e^{i\Phi_{EM}} & V_{EJ}e^{i\Phi_{EJ}} \\ V_{EM}e^{-i\Phi_{EM}} & E_M - i\gamma_M & 0 \\ V_{EJ}e^{-i\Phi_{EJ}} & 0 & E_J - i\gamma_J \end{pmatrix} \tag{2}$$

where the interaction between the effective $|ED\rangle$ contribution and the J-aggregate excitonic state $|J\rangle$ is described by the coupling parameter $V_{EJ}$ and the relative phase $\Phi_{EJ}$, which captures the phase relation of the near-field interaction (Figure 3a, ii). In this effective model, the added terms encode how the excitonic transition modifies the ED-MD-based chiroptical response of the helicoid. The coupling between $|MD\rangle$ and $|J\rangle$ is neglected because of the substantial detuning between the J-aggregate exciton and the MD resonance. The resulting modal decomposition, C1, C2 and C3, reproduces the split experimental lineshape very well (Figure 3c, Figure S11 and Table S2 in Supplementary Note 5) and yields two key insights. First, the coupling-induced C3 contribution is more consistent with an interference-driven, Fano-like response than with a conventional strong-coupling picture[53,54]. Quantitatively, the extracted interaction strength ($V_{EJ} = 42$ meV) is smaller than the scale set by the dissipative linewidths $(\gamma_{ED} - \gamma_J)/2 \approx 130$ meV, and therefore insufficient to meet the usual criterion for strong coupling, yet it remains well beyond a perturbatively weak interaction and is accompanied by distinct near-field signatures[55]. Together, these features place the system in an intermediate coupling regime in which spectral splitting appears in the far-field response (Figure S12 in Supplementary Note 5). We therefore use the terms UB and LB operationally to denote the higher- and lower-energy components of the hybrid optical response, rather than to imply fully developed polariton eigenstates in the strong-coupling limit. In this intermediate regime, the spectral splitting is shaped substantially by interference and dissipation, but the two energy-resolved components can nevertheless possess distinct local electromagnetic distributions. Second, even within this regime, the helicity dependence of the plasmonic basis response modifies the effective excitonic detuning and the local interaction profile. This is reflected in the helicity-dependent modal decomposition, which shows a deeper spectral dip in the C3 contribution under RCP than under LCP (Figure 3c).

The interference-driven features identified in the far-field spectra are directly reflected in the microscopic near-field response of the system, giving rise to polarization-dependent local hybridization. In the helicoid, each chiral motif constrains the surface currents and thereby modulates the local phase-correlated mixing of ED and MD moment densities, producing a pronounced circular contrast in the near-field (Figure S9 in Supplementary Note 4). Wave-optics simulations on the field intensity show that RCP illumination concentrates the field within the chiral gap, whereas LCP shifts it toward the particle periphery (Figure 3d; Supplementary Note 7). Because this selectivity is dictated by the chiral morphology of the helicoid, it is expected to reverse for the opposite helicoid handedness. The gap also governs the spatial character of the two energy-resolved hybrid components: the LB-like response is preferentially localized within the gap, while the UB-like response is distributed more strongly along the outer corners (Figure 3e). We attribute this asymmetry to the enhanced chiroptical activity originating from the gap at the LB wavelength (Figure 2e). In agreement with the simulations, loading-vector decomposition of the EEL spectra isolates the spatial distributions associated with these split spectral components, supporting the contrast between gap-localized LB-like response and corner-weighted UB-like response (Figure 3f; Figures S7 and S8 in Supplementary Note 3)[56].

Crucially, both the polarization-dependent near-field contrast and the spatial separation of the plexcitonic states are centered at the chiral gap, so their spatial profiles strongly overlap (Figure 3d and e). This combined spatial and spectral selectivity allows light polarization to act as an effective probe of specific hybrid states within localized regions of the particle. In addition, the chiral gap hosts a locally enhanced interaction between the plasmonic field and the excitonic J-aggregates. As a result, the helicity of light functions as a spectroscopic selector: RCP preferentially addresses the gap interior and its associated relaxation pathways, whereas LCP probes the particle periphery more strongly. Polarization-resolved TA spectroscopy is therefore well suited to resolve the resulting differences in ultrafast dynamics within the plexcitonic system.

**Polarization-resolved ultrafast dynamics of the helicoid-TDBC plexcitonic system**

To translate the localized spatial features of the coupling into the temporal domain, we employed a custom-built TA setup equipped with probe-pulse polarization control (Figure 4a, Supplementary Note 8). In the pristine helicoids, the time-resolved spectra exhibit a ground-state bleach at the plasmonic resonance, consistent with excitation of the plasmonic mode (Figure 4b)[57]. Upon conjugation with TDBC J-aggregates, this bleach feature becomes clearly split by the excitonic resonance, showing that the coupling signatures identified in the steady-state spectra translate into the ultrafast regime (Figure 4c). The kinetic traces further reveal that, in the pristine helicoids, the bleach recovery at the plasmonic maximum is essentially independent of probe polarization, with nearly identical dynamics for different probe helicities, indicating that the relaxation is spatially fairly uniform across the particle (Figures 4d). In the plexcitonic system, by contrast, a clear polarization dependence emerges. The UB, which is weighted towards the periphery, still shows only a weak dependence of the bleach recovery on probe-polarization (Figure 4e). The LB, however, which is mainly localized in the chiral gap, exhibits distinctly faster bleach recovery under RCP probing than under LCP probing. (Figure 4f). This is the central experimental observation of the time-resolved measurements: probe polarization selectively weights different nanoscale regions of the coupled system and thereby exposes their distinct ultrafast relaxation dynamics.

With this polarization-selective nanoscale probe in hand, we next consider the ultrafast relaxation pathways in the plexcitonic helicoids. Upon optical excitation, the coupled system accesses relaxation channels beyond those of bare plasmons, where the dynamics are typically governed by sub-picosecond electronic thermalization followed by lattice heating on the few-picosecond timescale[58,59]. In J-aggregates, optical excitation initially prepares the bright excitonic state, whereas the subsequent dynamics involve redistribution into a much larger manifold of optically dark states[60]. Although these dark states do not contribute directly to the far-field hybrid spectrum, their high density makes them an important relaxation reservoir[25]. In the present system, we propose that access to this dark-state manifold is enhanced for the gap-localized LB response. This is because the near-field in the chiral gap is not only strongly confined but also highly inhomogeneous on the molecular scale, so that excitonic states that remain dark under spatially uniform far-field excitation can couple more effectively to the local hybrid excitation. In this sense, the gap-localized near-field can open additional relaxation pathways into the dark-state manifold, analogous to a generalized Förster-type transfer from a bright donor excitation to nominally dark excitonic acceptor states. This picture is consistent with the differential dynamics observed in Figure 4d-f (see also Supplementary Note 9): under RCP probing, which preferentially addresses the chiral gap, the bleach recovery is faster because the gap-localized hybrid response relaxes more efficiently than the periphery-weighted response. To rationalize this behaviour, we formulated and solved a modified extended two-temperature model supplemented by bright- and dark-state excitonic channels (Supplementary Note 10)[61]. This phenomenological model reproduces substantially faster cooling in the gap region of the plexcitonic helicoid than in bare helicoid, with the contrast being particularly pronounced within the first 500 fs (Figures 4g, h), in agreement with the polarization-resolved TA measurements.

Ultimately, these findings show that plexcitonic coupling can be selectively interrogated across extreme spatiotemporal scales, spanning sub-wavelength spatial localization and sub-picosecond dynamics. A particularly clear illustration is provided by two synthetically tailored helicoids engineered to exhibit identical steady-state extinction under LP and RCP illumination, respectively (Supplementary Note 11). Despite their steady-state equivalence, the plexcitonic response probed under RCP exhibits significantly faster dynamics than that probed by LP. This pronounced kinetic contrast demonstrates that incident helicity selectively addresses distinct energy-redistribution pathways, allowing the nanoscale coupling landscape to be resolved in the temporal domain via far-field polarization.

## Conclusion

Taken together, our results show how the multiscale chirality of gold helicoids is manifested as spatial, temporal, and energetic asymmetry in plasmon-exciton coupling. By conformally functionalizing the helicoids with an excitonic layer, we translate morphologically encoded ED-MD interference into experimentally accessible observables and show that localized plexcitonic responses can be selectively addressed through light helicity. To the best of our knowledge, this work provides the first spatiotemporal mapping of chiral plexcitonic dynamics and establishes a physically grounded, component-resolved picture of the underlying hybrid response. Beyond the specific helicoid-J-aggregate system studied here, the combined use of helicity-resolved far-field spectroscopy, nanoscale spatial mapping, and polarization-resolved ultrafast dynamics establishes a general route for accessing spatially distinct hybrid states and their relaxation pathways in chiral nanophotonic systems[62-64]. Although demonstrated here using molecular J-aggregates, the underlying principles should be extendable to a wide range of organic and inorganic quantum emitters[65,66], as well as to more complex plasmonic assemblies and crystal architectures[67,68]. In this broader context, our findings point toward helicity-selective control of hybrid energy flow at the nanoscale and open opportunities for coupling chiral plasmonic structures with intrinsically chiral excitonic materials for future spin- and quantum-photonic functionalities.

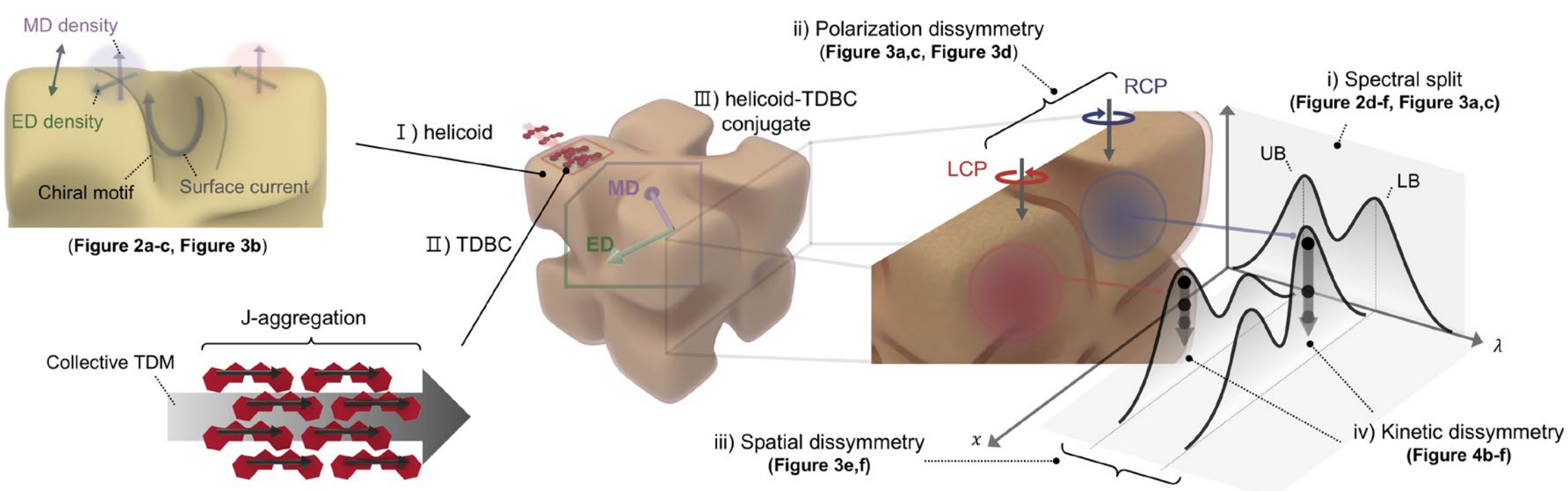

**Figure 1. Rational design of helicoid-TDBC plexcitonic system.** Schematic illustration detailing the central operating mechanisms of the individual constituents and their synergistic integration into a helicoid-TDBC plexcitonic system. In the core plasmonic helicoid (I), surface currents morphologically confined within the chiral gap induce a strong mixing of ED and MD moments and their local densities, acting as the fundamental origin of chiroptical responses. TDBC molecules (II) assemble into J-aggregates at the surface of helicoid, exhibiting a strong collective transition dipole moment (TDM) benefiting efficient light-matter interaction. Upon their conjugation (III), the coupling explicitly translates into four distinct dissymmetries: (i) Spectral splitting of the hybrid optical response into lower- and higher-energy components, denoted LB and UB; (ii) Helicity-driven spatial dissymmetry, selectively addressing the local density of states via circularly-polarized incidence; (iii) Interlevel spatial dissymmetry between UB and LB, enabling helicity-steered level-selective probing; and (iv) Kinetic dissymmetry, translating such spatial segregation into pathway-specific ultrafast relaxation dynamics.

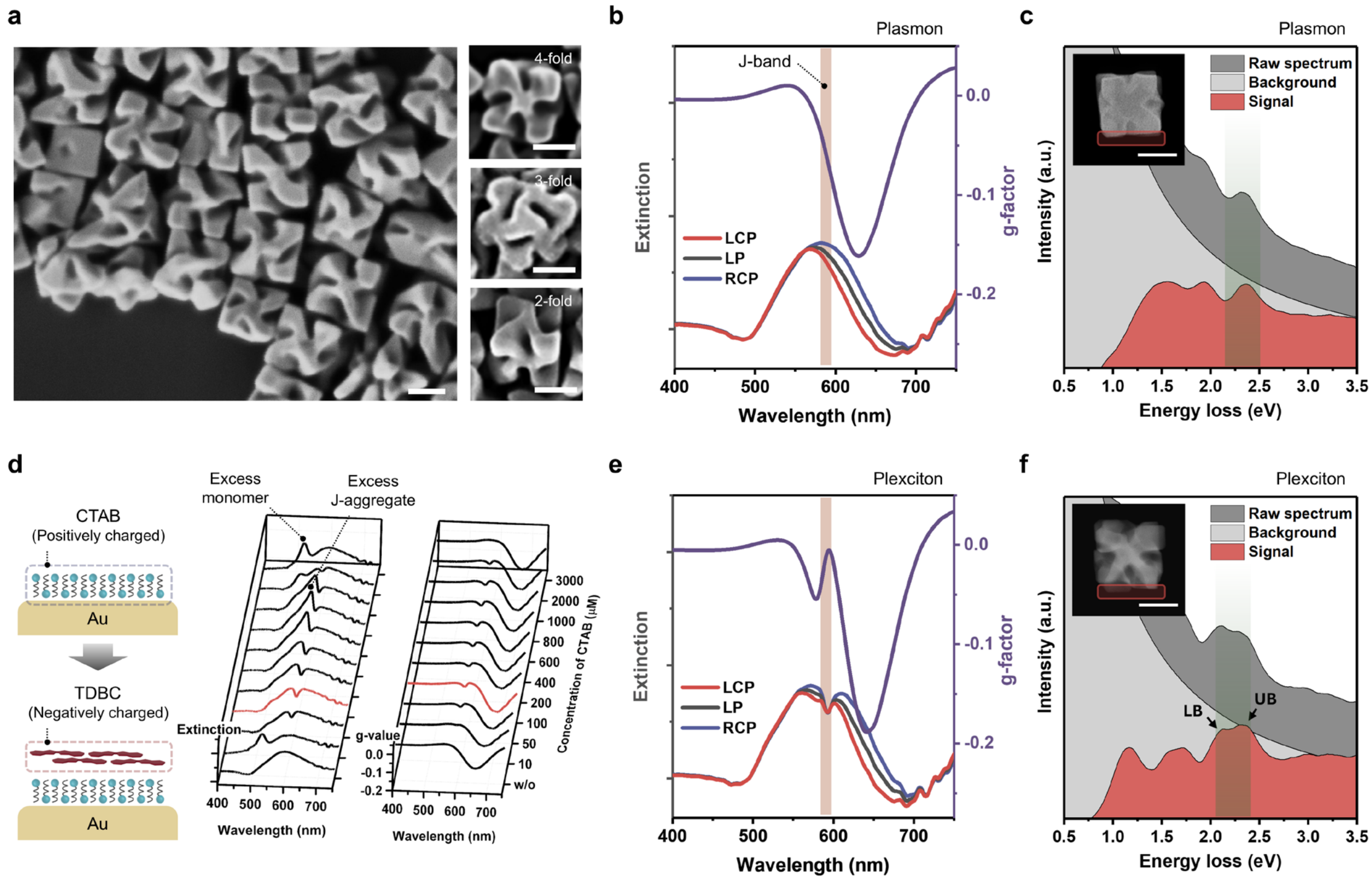

**Figure 2. Realization of helicoid-TDBC plexcitonic system. a**, SEM images of helicoids, revealing distinct three-dimensional chiral gap morphologies explicitly governed by 432 rotational symmetry. **b**, Steady-state chiroptical spectra of the pristine helicoids, where incident polarization actively dictates the plasmonic bands' peak wavelength, linewidth, and detuning with excitonic band. **c**, Electron energy loss spectroscopy (EELS) profile across the vicinal space of a pristine (plasmonic) helicoid. Subtracting the power-law fitted background confirms a distinct localized plasmon band at 2.3 eV (green shaded area), primed for overlap with TDBC J-aggregate's excitonic band. **d**, Electrostatic assembly of TDBC onto CTAB-capped helicoids investigated via CTAB concentration-dependent optical spectra. A concentration-dependent competitive interplay between surface passivation and self-micellization of CTAB systematically modulates the assembling and coupling features of TDBC. **e**, Steady-state chiroptical spectra of the plexcitonic helicoids, where the polarization-driven variations in the splitting magnitude and profile explicitly imply a helicity-governed coupling nature. **f**, EELS profile of the plexcitonic helicoid, directly capturing the spectral splitting at the 2.3 eV (green shaded area) band. Scale bars, 100 nm.

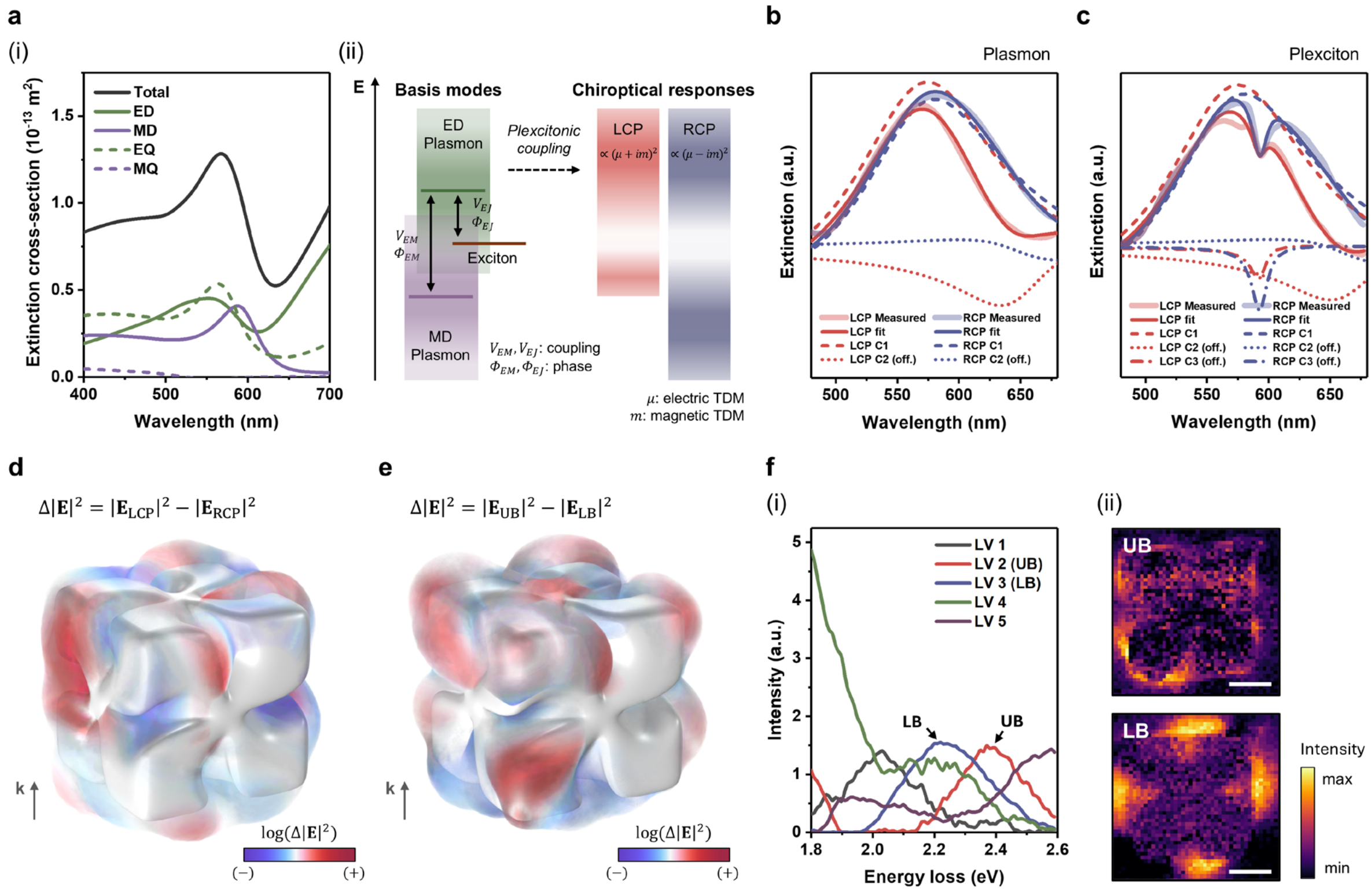

**Figure 3. Macro-to-microscopic characterization of helicoid-TDBC plexcitonic coupling. a**, Conceptual framework of the chiral plexcitonic interactions. (i) Multipole decomposition of the plasmonic helicoid, showing the concurrent excitation of ED and MD modes, which serves as the fundamental origin of its plasmonic chirality. (ii) Schematic illustrating the interference picture between the plasmonic continuum states (ED and MD) and the discrete excitonic states, leading to their characteristic optical attenuation under circularly polarized incidence. **b, c**, Extinction spectra (solid lines) and their modal contributions (dashed lines, C1 to C3) for the (**b**) plasmonic and (**c**) the plexcitonic helicoids, analytically retrieved via a non-Hermitian Hamiltonian model. **d**, Simulated circular differential near-field intensity profile ($|\mathbf{E}_{\mathrm{LCP}}|^2 - |\mathbf{E}_{\mathrm{RCP}}|^2$) of the plexcitonic helicoid at the LB energy, showing preferential field funnelling into the gap under RCP illumination. **e**, Simulated near-field intensity contrast between the UB and LB components. The LB field is strongly localized within the gap, whereas the UB distributes toward the periphery. **f**, (i) Loading vector decomposition of EEL spectra and (ii) the corresponding spatial field distributions. The field is tightly confined within the gap at the LB, while localizing at the outer edges at the UB.

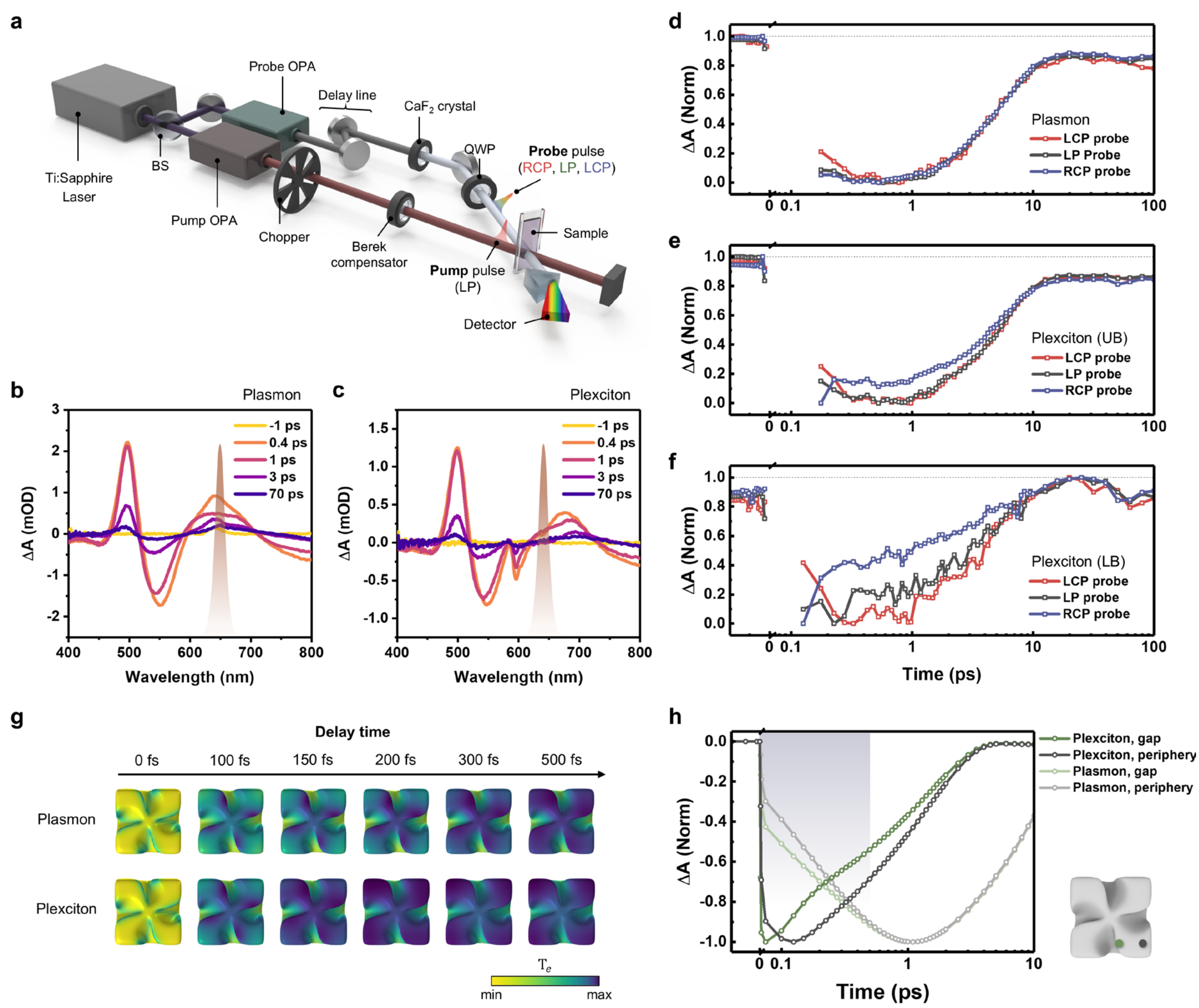

**Figure 4. Resolving and controlling ultrafast dynamics of helicoid-TDBC plexcitonic system. a**, Schematic illustration of the custom-built transient absorption (TA) spectroscopy setup with a polarization-controlled probe pulse. **b, c**, Slices of TA spectra for (**b**) plasmonic and (**c**) plexcitonic helicoids, following excitation at 640 nm (pumped and probed with linear polarizations). **d-f**, Ultrafast decay kinetics of plasmonic (**d**, 550 nm), UB (**e**, 550 nm) and LB (**f**, 590 nm) states of helicoids, pumped with 640 nm linear polarization and probed with different polarizations. An accelerated sub-picosecond plexcitonic decay appears under RCP probing, most prominently at the LB level, while there is almost no polarization dependence in plasmonic helicoids. **g**, Simulated electron temperature profiles in plasmonic and plexcitonic helicoids as a function of time, visually demonstrating the accelerated thermal relaxation of gap-localized hot electron distributions under the plexcitonic coupling. **h**, Simulated dynamics of normalized electron temperature at gap (green) and non-gap (black) region in plasmonic and plexcitonic helicoids from equilibrium, quantitatively resolving the gap-localized rapid plexcitonic decay. The blue-shaded region corresponds to the time window visualized in **g**.

**Supplementary information** is available in the online version of the paper.

**Acknowledgments** This research was funded by the European Union under HE-GA 101131771 Lasers4EU (project ID: 27885). This research was also supported by the National Research Foundation of Korea (NRF) grant funded by the Korea government (MSIT) (RS-2024-00409405, RS-2024-00436855). T.P. acknowledges support by Swedish Foundation of Strategic Research (IS24-0005), Olle Engkvists Foundation (235-0422) and KAW WACQT program. K.T.N. appreciates the administrative and technical support from Institute of Engineering Research, Research Institute of Advanced Materials (RIAM), and SOFT foundry institute at Seoul National University.

**Author contributions:** J.H.H. and S.R. contributed equally. T.P. and K.T.N. conceived and supervised this work. Material design and synthesis was conducted by J.H.H., S.R., R.M.K., I.H.H., and J.L. Steady-state spectroscopic characterization was carried out by J.H.H., S.R., R.M.K., and S.H.C. Non-Hermitian Hamiltonian modelling was performed by J.H.H., S.R., and A.S. Ultrafast spectroscopic characterization was conducted by J.H.H., S.R., P.C., and R.M.K. TEM characterization was carried out by J.J., Y.T., and M.K. Numerical simulation was performed by J.H.H. and S.R. J.H.H., S.R., T.P., and K.T.N. wrote the manuscript. All authors discussed the results and commented on the manuscript.

**Author Information** The authors declared no competing financial interests. Correspondence and requests for materials should be addressed to T.P. (tonu.pullerits@chemphys.lu.se) and K.T.N. (nkitae@snu.ac.kr).

## Methods

**Chemicals** Hexadecyltrimethylammonium bromide (CTAB, 99%), L-ascorbic acid (AA, 99%), Cetyltrimethylammonium chloride solution (CTAC, 25 wt. % in $H_2O$), tetrachloroauric (III) trihydrate ($HAuCl_4 \cdot 3H_2O$, 99.9%), L-glutathione (L-GSH, 98%), sodium borohydride ($NaBH_4$, 99%) were purchased from Sigma-Aldrich. 5,6-Dichloro-2-[[5,6-dichloro-1-ethyl-3-(4-sulfobutyl)-benzimidazol-2-ylidene]-propenyl]-1-ethyl-3-(4-sulfobutyl)-benzimidazolium hydroxide, inner salt, sodium salt (TDBC) was purchased from Few Chemicals. All aqueous solutions were prepared using high-purity deionized water (18.2 MΩ $cm^{-1}$).

**Synthesis of Helicoids** First, Helicoid nanoparticles were synthesized through a peptide-directed, seed-mediated method. The growth solution was prepared by adding 800 µL of CTAB (100 mM), 0.2 µL of $HAuCl_4$ (10 mM), 475 µL of ascorbic acid (100 mM), and 5 µL of L-GSH (5 mM) to 3.95 mL of deionized water. Then, 50 µL of octahedral seed nanoparticles, prepared by previously reported methods[46], was injected into the growth solution and the mixture was left undisturbed at 30 °C for 2 h. Fully grown colloids were centrifuged twice (2000 × g, 3 min) to remove residual reagents and finally redispersed in 1 mM CTAB to have 2 OD at 400 nm, 10 mm pathlength.

**J-aggregate conjugation on Helicoids** J-aggregation of TDBC molecules on helicoid nanoparticles was achieved using a previously reported conjugation protocol. The process relies on electrostatic self-assembly of negatively charged TDBC molecules onto the positively charged CTAB ligands surrounding the helicoid surface. To avoid spectral obscuration by excess TDBC and competitive assembly on CTAB micelles, the relative concentrations were carefully optimized (Supplementary Note 2). Prior to conjugation, the helicoid solution was centrifuged and redispersed in a CTAB solution with a reduced concentration (100 µM). The dispersion was then mixed under sonication with an equal volume of an aqueous TDBC solution (10 µM) and kept in the dark for 5 min.

**Steady-state optical characterization** Circular dichroism and extinction were measured using J-1700 CD Spectrometer (JASCO). Kuhn's dissymmetry factor *g* was calculated from the measured CD and extinction:

$$g = 2\frac{A_{LCP} - A_{RCP}}{A_{LCP} + A_{RCP}} \propto \frac{CD}{extinction}$$

**Transient absorption spectroscopy** TA measurements were conducted by using a home-built femtosecond pump-probe setup described in earlier work[10,69]. Solstice (Spectra Physics) amplifier seeded by a femtosecond oscillator (Mai Tai SP, Spectra Physics) provides laser pulses (8 W, 796 nm, 60 fs, 4 kHz). It is split into two beams that each pump collinear optical parametric amplifiers (TOPAS-C, Light Conversion). The first one generates 640 nm wavelength pump pulses, while the other generates 1350 nm pulses that is focused onto a $CaF_2$ plate to generate a supercontinuum pulse which is used as the probe. The probe beam is delayed with respect to the pump using a mechanical delay stage. After supercontinuum generation, the probe beam is further divided into two parts: the former is focused on the sample overlapping with the pump pulse and the latter serving as a reference. After passing the sample, the probe beam is collimated again and relayed onto the entrance aperture of a prism spectrograph. The reference beam is directly relayed on the spectrograph. Both the measurement and reference beams are then dispersed onto a double photodiode array, each holding 512 elements (Pascher Instruments). The pump beam is made linearly polarized, perpendicular to the probe polarization with a Berek compensator working as a half wave plate. This configuration allows to reduce excitation light scattering from reaching the spectrograph by placing a Glan-Thompson polarizer after the sample and before the spectrograph.

For TA with circular-polarized pulses, the above-mentioned setup is modified with additional optics, described in earlier work. Identical achromatic quarter wave-plates (Thorlabs AQWP05M-600, 400-800 nm) are placed in the path of probe beam before the sample. The angles of the wave-plate required for circular polarization were determined using Stokes-polarimetry based polarimeter (Thorlabs PAX1000VIS/M). The experiments were performed for a pump laser power ranging from 1 to 13.3 mW. Data analysis of TA experiments was performed with Origin 2025 and custom-written scripts in python.

**Numerical simulations** Numerical simulations were carried out in COMSOL Multiphysics, commercial finite-element-method (FEM) based Maxwell equation solver. The helicoid geometry was represented by a properly scaled three-dimensional model consistent with that used in previous work[34], and the optical constants of gold were taken from the Johnson and Christy dataset available in the COMSOL material library. Dispersion of optical constant of excitonic TDBC J-aggregate was modeled based on Lorentz oscillator model (Supplementary Note 7)[27], and the surrounding medium was assigned a refractive index of water, 1.33.

The steady-state optical response was retrieved solely by solving the scattering problem in COMSOL in the wavelength domain. Subsequently, the spatiotemporal dynamics were calculated by modeling the temporal dissipation of the optical gain, derived from the steady-state response, within the plasmon-based system. An extended two-temperature model (eTTM) governing the intrinsic plasmonic dynamics, and two partial differential equations for the excitonic bright state and dark state that incorporate intercorrelation terms, were solved simultaneously[61]. This coupled system was initialized by allocating the initial thermal energy distribution retrieved from the wave optics simulations (Supplementary Note 10).

**Electron Energy Loss Spectroscopy** The helicoids after J-aggregate conjugations were dispersed on a copper grid with an ultrathin carbon supporting film (Ted Pella, USA) for EELS analysis. EELS spectrum images were acquired using a 300-keV monochromated Cs-corrected TEM (Themis Z, Thermo Fisher Scientific, USA) equipped with GIF Quantum ERS (Gatan Inc. Ametek, USA). The convergence and collection semi-angle were set as 3.65 mrad and 19.03 mrad, respectively. Each pixel spectrum consisted of 2,048 energy channels with an energy dispersion of 0.01 eV per channel and a dwell time of 0.03 s. The Full-width-half-maximum of the zero-loss peak on the support film was around 100 meV, so that the spectral splitting of plexcitonic helicoid system (100 meV) was able to be reliably resolved.

Supplementary Information

# Helicity-Resolved Spatiotemporal Mapping of Chiral Plexcitons in Helicoids

*Jeong Hyun Han[1†], Sankaran Ramesh[2†], Jaeyeon Jo[1,3], Pavel Chabera[2,4], Ryeong Myeong Kim[1,5], Sung Hoon Cho[1], In Han Ha[1], Amitav Sahu[2], Yoonsang Tak[1], Jiawei Lv[1], Miyoung Kim[1,3], Ki Tae Nam[1*], Tönu Pullerits[2*]*

[1] Department of Materials Science and Engineering, Seoul National University, Seoul 08826, Republic of Korea

[2] Division of Chemical Physics and NanoLund, Lund University, 22100 Lund, Sweden

[3] Research Institute of Advanced Materials, Seoul National University, Seoul 08826, Republic of Korea

[4] Department of Physics, College of Science, United Arab Emirates University, P.O. Box 15551, Al Ain, United Arab Emirates

[5] Emerging Material Metrology Group, Division of Chemical and Material Metrology, Korea Research Institute of Standards and Science (KRISS), Daejeon 34113, Republic of Korea

**† These authors contributed equally to this work.**

*** To whom correspondence should be addressed:**

nkitae@snu.ac.kr, tonu.pullerits@chemphys.lu.se

## Table of Contents

## Supplementary Note 1. Steady-state optical characterization of helicoids and TDBC

TDBC (5,6-Dichloro-2-[[5,6-dichloro-1-ethyl-3-(4-sulfobutyl)-benzimidazol-2-ylidene]-propenyl]-1-ethyl-3-(4-sulfobutyl)-benzimidazolium hydroxide) can exist in either monomeric or J-aggregated states in aqueous solution, depending on concentration, ionic environment, and the presence of assembly sites. Upon J-aggregation, the coherent coupling of individual transition dipole moments (TDMs) yields a collective TDM, which provides enhanced resistance to dephasing. This makes TDBC a highly suitable candidate for realizing plasmon-exciton coupled states. In an aqueous environment, the aggregation state of TDBC—and consequently its optical properties—can be precisely tuned simply by varying the concentration. At a low concentration of 5 μM, the spectrum is dominated by a monomeric absorption band at 518 nm; however, as the concentration is increased to 50 μM, an excitonic band at 590 nm, characteristic of J-aggregation, prominently emerges (Figure S1a). Furthermore, the absence of intrinsic circular dichroism (CD) at these concentrations ensures that the plexcitonic chirality observed in our system is exclusively governed by the plasmonic chirality of the helicoid (Figure S1b).

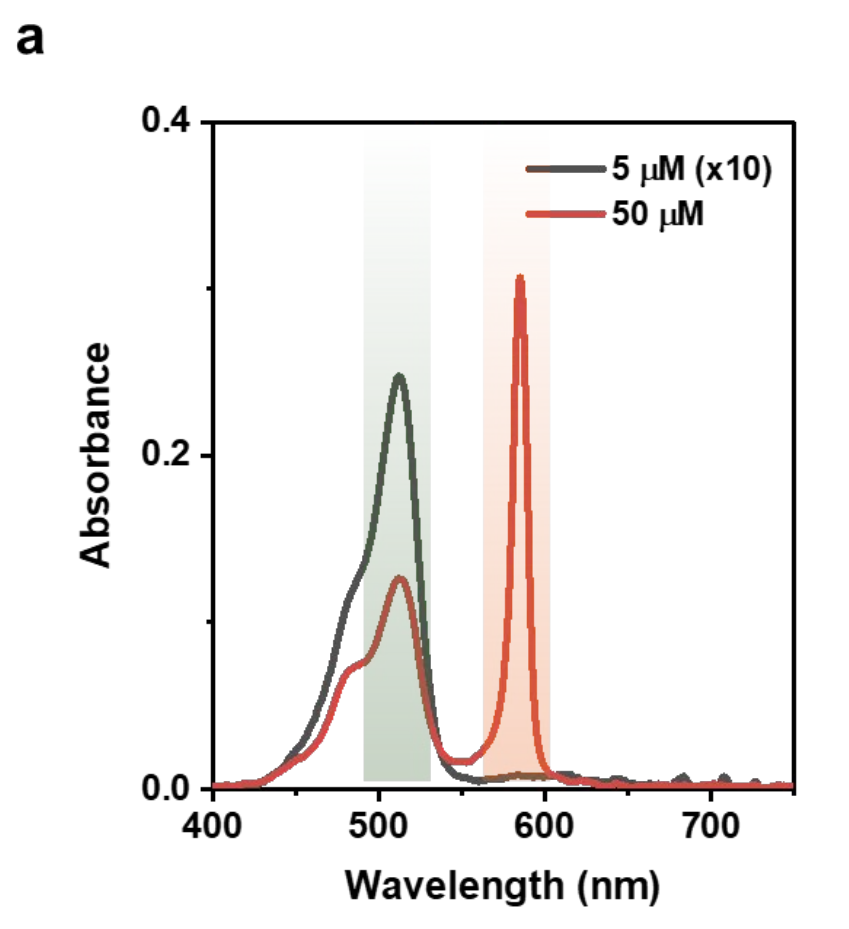

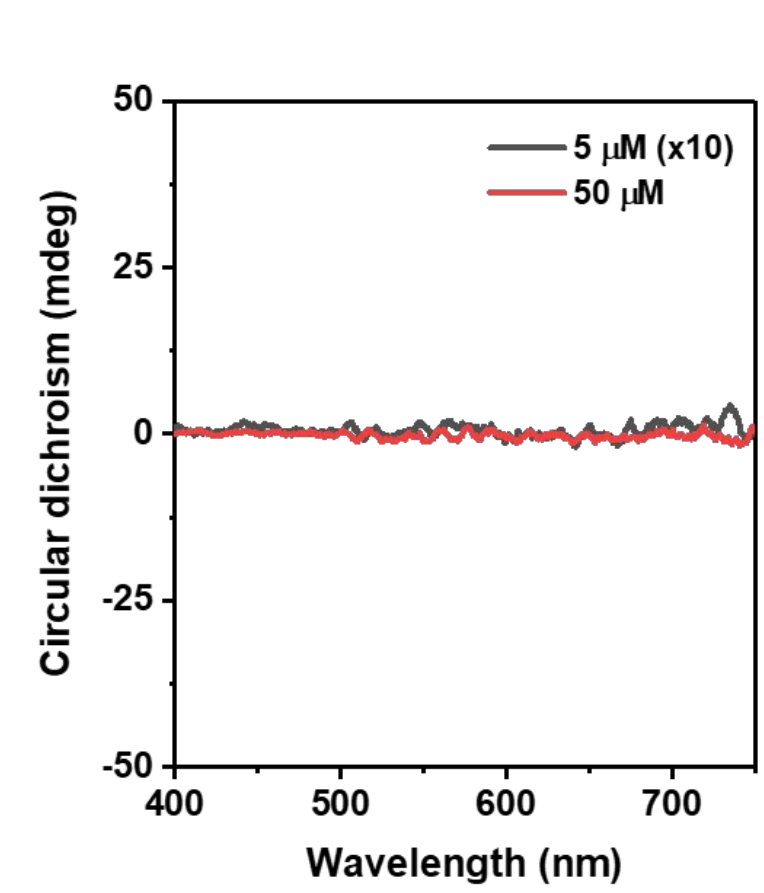

**Figure S1. Steady-state optical characterization of TDBC. a,** Absorbance spectra of TDBC aqueous solutions at various concentrations. At low concentration (5 μM), the monomeric absorption band (green-shaded region) at 518 nm dominates, whereas the excitonic band (orange-shaded region) at 590 nm arising from J-aggregation emerges as the concentration increases (50 μM). **b,** CD spectra of the TDBC solutions at corresponding concentrations. Neither the monomeric nor the J-aggregated states exhibit significant intrinsic CD. 5 μM measured in ×10 pathlength compared to 50 μM.

Two variants of helicoids, denoted as Helicoid$_{(A)}$ and Helicoid$_{(B)}$ were investigated in this study, with their extinction and CD bands precisely tuned via a stoichiometry in seed-mediated synthesis step[1]. Two distinct configurations were established to meet specific resonance conditions: (1) Helicoid$_{(A)}$ was designed such that its LSPR band under RCP incidence exhibits minimal detuning with the TDBC excitonic band, with the maximum absolute *g*of the coupled system positioned at the 640 nm pump wavelength for direct LB excitation (640 nm) (Figure S2a, also featured in Figures 2b and 2e). (2) Helicoid$_{(B)}$ was engineered to exhibit an identical spectral profile under unpolarized (or linearly polarized) light as Helicoid$_{(A)}$ does under RCP, minimizing detuning between the LSPR and excitonic levels (Figure S2b, also featured in Figures 2d). This strategic design allows for the independent isolation of polarization-dependent near-field coupling characters even in far-field optical measurements (see Supplementary Note 11).

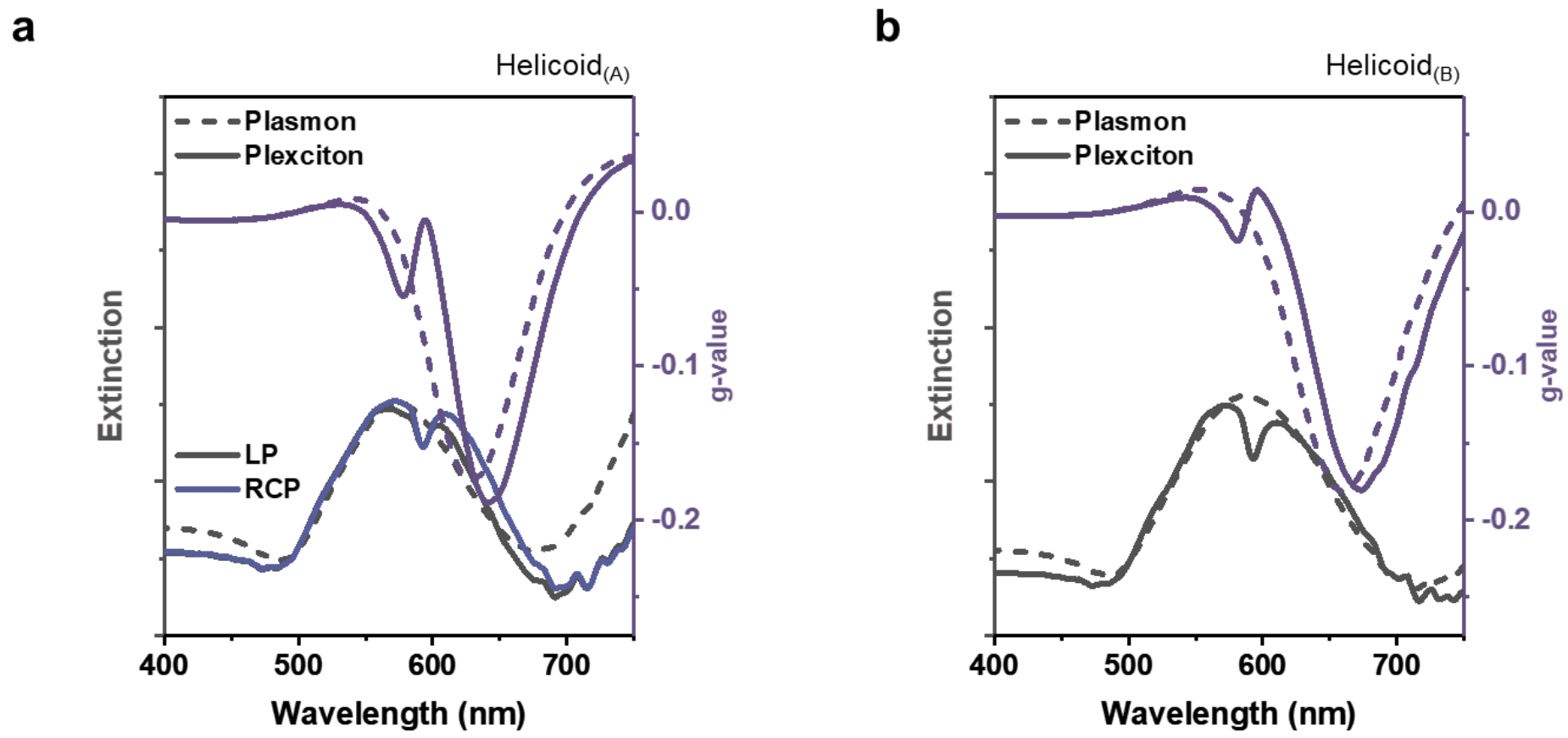

**Figure S2. Steady-state optical characterization of Helicoid$_{(A)}$ and Helicoid$_{(B)}$. a, b,** Extinction and CD spectra of (**a**) Helicoid$_{(A)}$ and (**b**) Helicoid$_{(B)}$. The two variants are synthetically designed such that the RCP extinction profile of Helicoid$_{(A)}$ exactly matches the LP extinction profile of Helicoid$_{(B)}$.

## Supplementary Note 2. Engineered conjugation of helicoids and TDBC

The conjugation of TDBC onto helicoids leverages the electrostatic assembly between negatively charged TDBC molecules and the positively charged surface ligands (CTAB) of the helicoids. This approach ensures conformal and total surface coverage, which is essential for precisely harnessing the sub-diffraction-limited morphological features of our system. Achieving an optimized coupling requires a mechanistic balance between CTAB and TDBC, as their interaction follows three distinct regimes: (1) At low CTAB concentrations (< 100 μM), incomplete surface coverage results in uncoupled TDBC monomers; (2) At intermediate conditions (100 μM - 1 mM), absorption of excess floating J-aggregates becomes noticeable, potentially masking the coupling signatures; and (3) Above the critical micelle concentration (CMC, > 1 mM), CTAB micelles isolate TDBC molecules, disrupting J-aggregation (Figure S3), triggering the emergence of monomeric absorption band[2]. Considering our experimental workflow—where spectroscopic measurement including transient absorption (TA) necessitates reproducible and reliable optical density control—this precisely tuned one-pot synthesis offers a significant advantage.

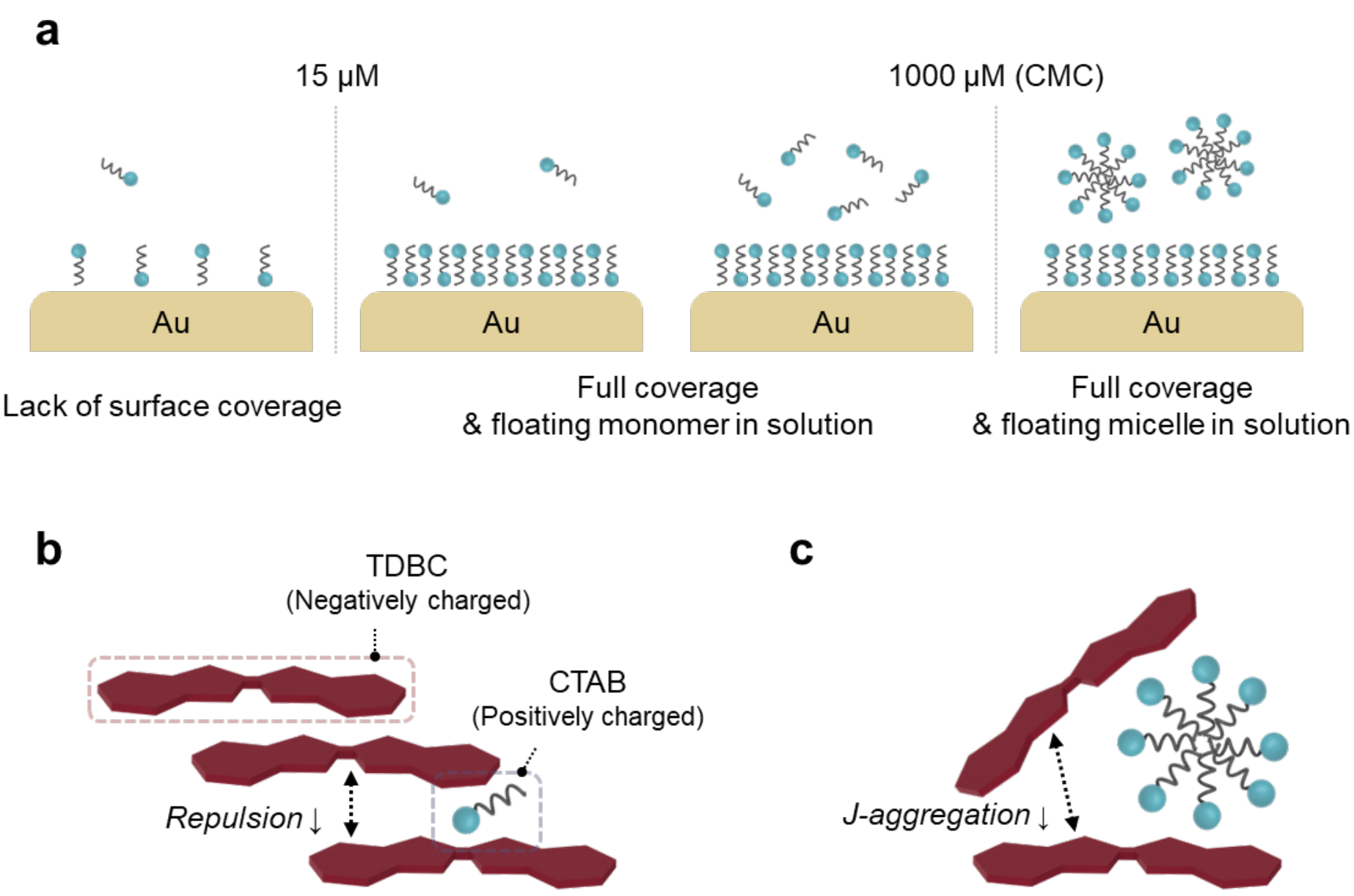

**Figure S3. Schematic illustration depicting the role of CTAB in the aggregation of TDBC. a,** CTAB concentration-dependent surface coverage and the corresponding morphological states of gold helicoids in aqueous solution. **b,** Reduction of intermolecular repulsion between negatively charged TDBC molecules mediated by free $CTA^{+}$ ions during J-aggregation. Below the critical micelle concentration (CMC), this promotes the formation of freely floating excitonic J-aggregate species rather than the conjugation onto the helicoid surface. **c,** Hindrance of J-aggregation induced by micelle formation, an effect that becomes dominant at CTAB concentrations above the CMC.

While increasing the TDBC concentration ($N$) can enhance collective coupling ($g \propto \sqrt{N}$), an overabundance in our one-pot scheme introduces excess floating J-aggregates that not only obscure the spectral split features but also contaminate the ultrafast dynamics. Consequently, by fixing the CTAB concentration at an optimized 100 μM, we identified 10 μM of TDBC as the ideal threshold to maximize the spectral signature of the plexcitonic split while minimizing background interference (Figure S4).

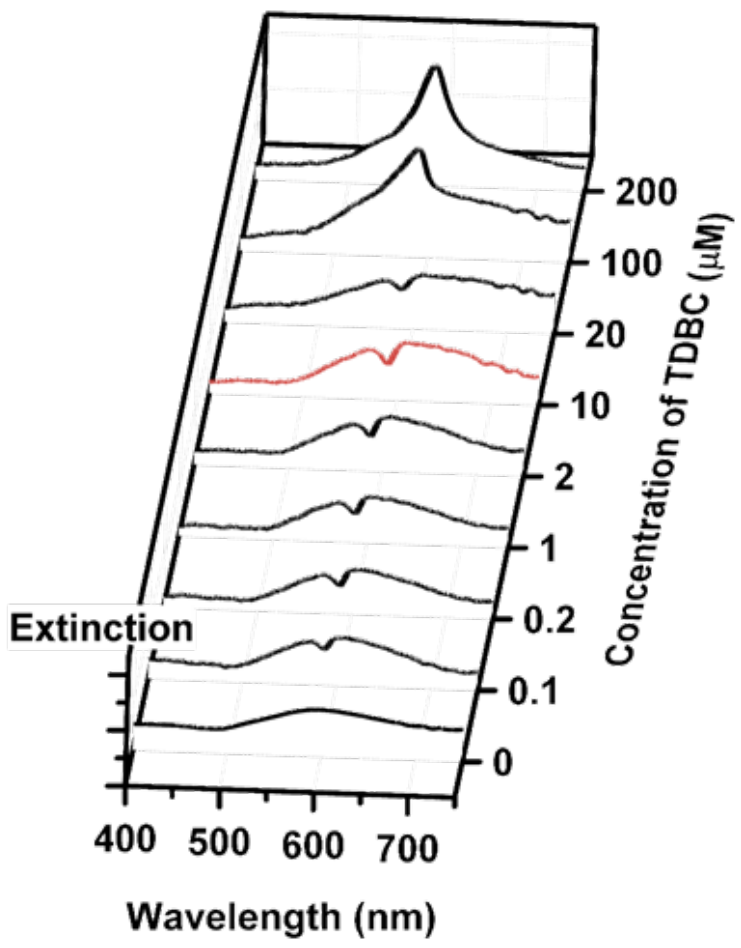

**Figure S4. TDBC concentration-dependent extinction spectra of helicoid-TDBC conjugate.** Increasing the TDBC concentration up to 10 μM strengthens the coupling constant ($g$) and produces a more apparent spectral split. Beyond this concentration, the splitting starts to be masked by the contribution of excess J-aggregates residing outside the conjugation.

Under these precisely engineered conditions, elemental mapping using transmission electron microscopy energy-dispersive X-ray spectroscopy (TEM-EDX) confirms the successful conjugation, revealing a conformal distribution of sulfur (S) and chlorine (Cl), elements exclusive to the TDBC molecules, across the helicoid surface (Figure S5).

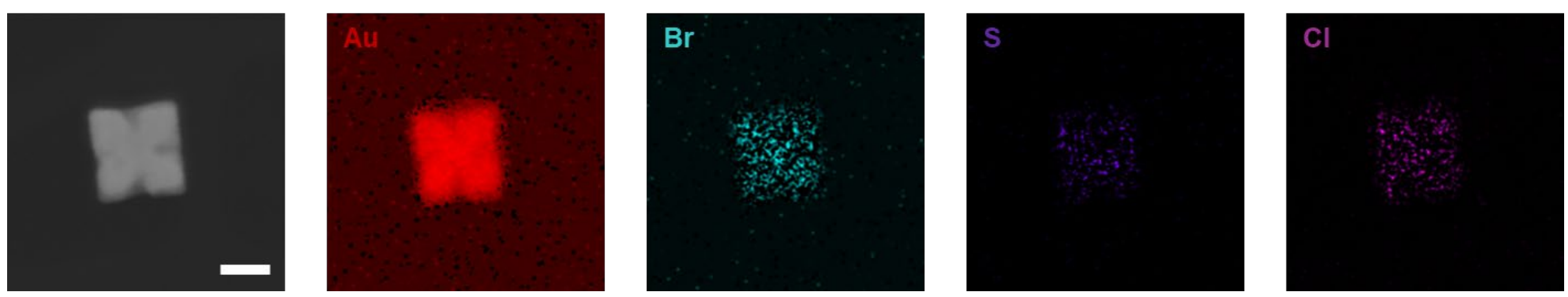

**Figure S5. TEM-EDX analysis of the helicoid-TDBC conjugate.** The elemental mapping was acquired over the corresponding HAADF-STEM image. Scale bar, 100 nm.

## Supplementary Note 3. EELS analysis on plexcitonic helicoids

In our work, Electron Energy Loss Spectroscopy (EELS) serves as a powerful analytical tool that extends beyond far-field optical spectroscopy and wave-optics numerical simulations. The electron beam provides a probe capable of decomposing intrinsic modes with spatial resolution far beyond the optical diffraction limit. Thereby, EELS enables experimental confirmation and characterization of local plasmon-exciton coupling. This offers a physically robust and unique perspective on the spatial distribution of the hybridized plexiton states that is inaccessible through conventional optical means.

Firstly, the split signatures of the EEL spectra, collected by varying the beam position around the plexcitonic helicoid, reflect the spatially dependent nature of the plexcitonic coupling in helicoid (Figure S6). Notably, the splitting becomes more pronounced as the gap-localized plasmon mode intensifies, indicating a locally enhanced coupling at the chiral gap that induces a significant spatial asymmetry. We anticipate that this spatial contrast provides collateral evidence for an interference-governed, intermediate-coupling nature as identified by the NHH model (see Supplementary Note 5), rather than the fully coherent eigenmode hybridization typically expected in the strong-coupling regime.

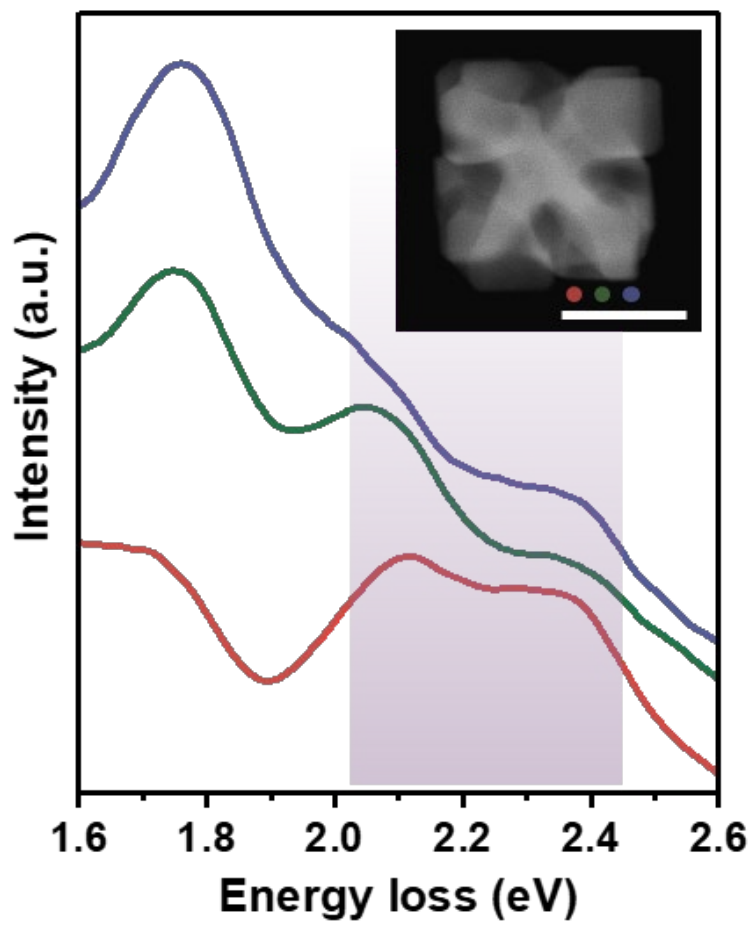

**Figure S6. EEL spectra of helicoid-TDBC conjugate probed at varying electron beam position.** Beam position indicated with red, green, and blue circles in the inset HAADF image. Purple-shaded region corresponds to the energy region of interest, where the plexcitonic states due to the coupling between helicoid and TDBC J-aggregate are visible in the absorption spectrum

We employ a dimensionality reduction (DR) approach to effectively decouple the modal contributions from the two-dimensional EELS dataset, which is inherently high-dimensional since every individual spatial pixel contains a full EEL spectrum comprising hundreds of energy channels. This specific methodology was adopted from previous reports that demonstrated effective DR processing for helicoid nanoparticles[3]. Briefly, non-negative matrix factorization simplifies the dataset by approximating the raw spectra as strictly additive combinations of isolated signals. By decomposing the original data into a few basic spectral components (*i.e.*, loading vectors, LVs) and their corresponding spatial weights (*i.e.*, coefficient maps), it achieves this reduction while maintaining strict physical interpretability through non-negativity constraints. This data-driven reduction allows us to isolate distinct local plasmonic excitations without requiring prior physical models. It is particularly advantageous for our system because it provides phenomenological information on modal contributions without imposing biased physical constraints on the coupled regime. We also expect that this technique largely excludes potential background sources other than the plexcitonic states. Consequently, measured consistently across multiple particles, the LV spectra and their mapped spatial distributions clearly reflect the spatial asymmetry between the corner-distributed UB and the gap-localized LB (Figure 3f and Figure S7).

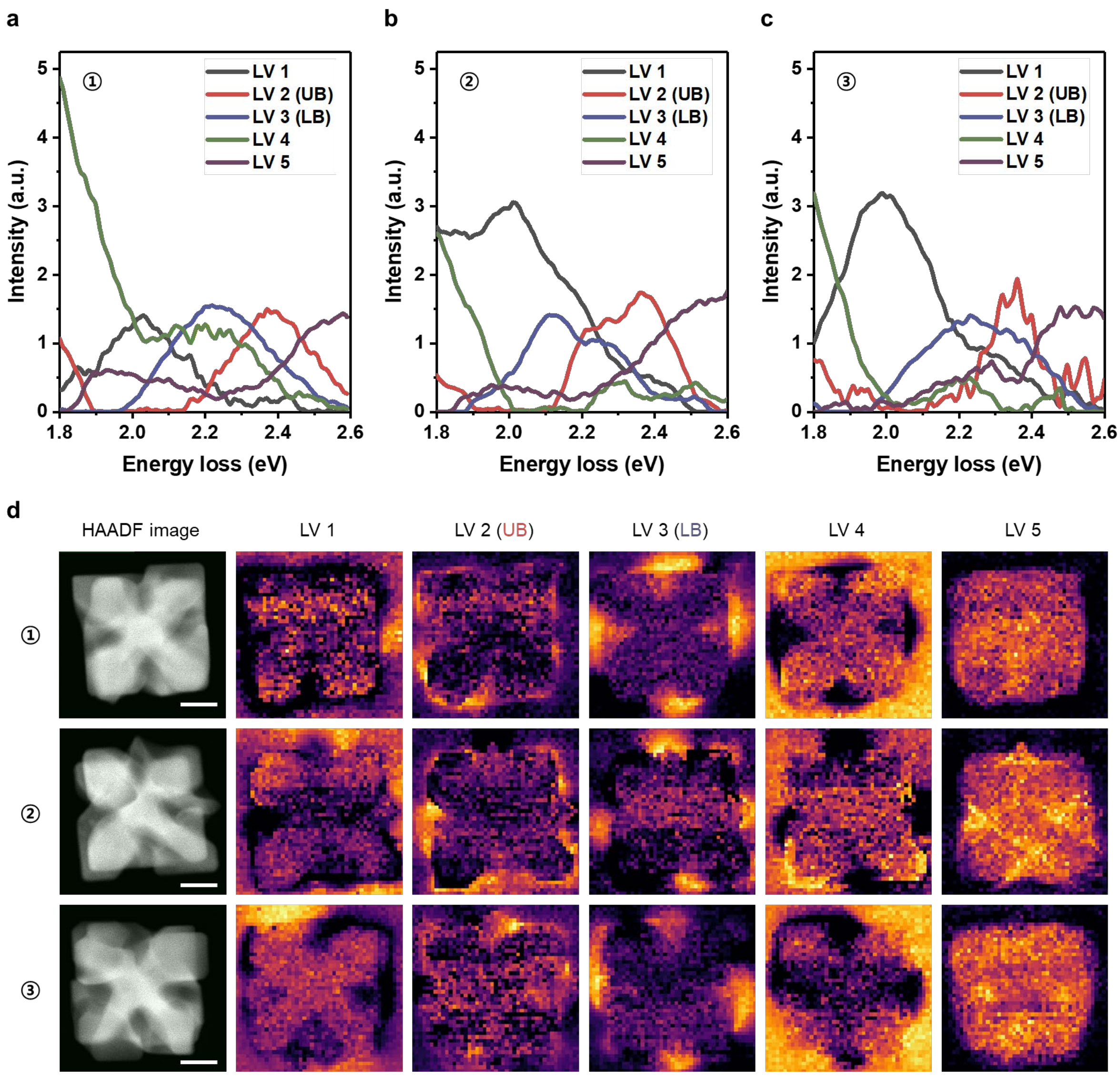

**Figure S7. Dimensionality reduction analysis of EELS datasets acquired from multiple helicoid-TDBC conjugates. a-c,** Loading vectors and **d,** their corresponding coefficient maps for three different helicoid-TDBC conjugates, consistently revealing UB localization at the corners and preferential LB focusing within the gaps. Scale bars, 50 nm.

To establish direct correspondence with the optical regime, we line-integrated the simulated 3D field distributions of the UB and the LB states (featured in Figure 3e) along the incident k-vector, thereby emulating the measurement physics of EELS (*i.e.*, the integration of energy loss along the electron's trajectory) (Figure S8, simulation details in Supplementary Note 7). The resulting profiles cross-validate that the spatial dissymmetry between the system is an intrinsic property of the system, fundamentally independent of the excitation or probing modality.

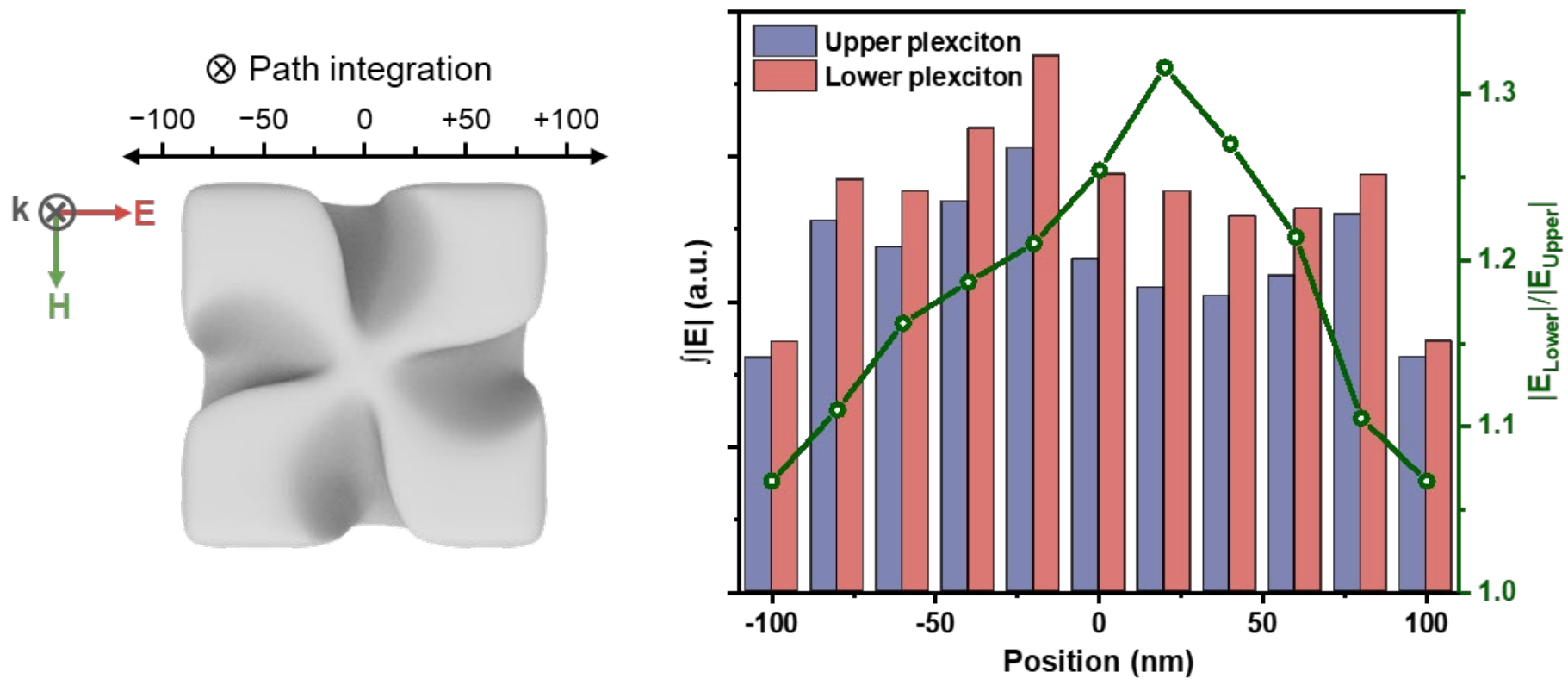

**Figure S8. Simulation for the line-integrated optically excited near-field of helicoid-TDBC conjugate.** Comparison of the scattered electric fields at the UB and LB levels, integrated along the k-vector direction as a function of lateral position.

## Supplementary Note 4. Numerical analysis on ED and MD interferences in helicoids

In order to study the multipolar modal contribution in the plasmonic resonance of helicoids, we performed and analysed three-dimensional full-wave electromagnetic simulations using the wave optics module in COMSOL Multiphysics, finite-element-method (FEM) based commercial Maxwell equation solver. The computational domain was defined as a spherical core-shell geometry, in which a perfectly matched layer (PML) surrounded the calculation region to absorb residual electromagnetic waves. The gold helicoid three-dimensionally modelled in previous report was positioned at the centre of the domain, and the complex total electric field $\mathbf{E}(\mathbf{r})$ under illumination of a monochromatic plane wave was obtained in the wavelength domain. The dielectric parameters $\varepsilon_r$ of the gold were taken from the previous report by Johnson and Christy available in the material database of COMSOL Multiphysics while the surrounding medium was treated as a homogeneous dielectric host with dielectric constant $\varepsilon_h$ of water.

To interpret the optical response within the current-based multipole formalism, the simulated internal field was converted into the surface current density $\mathbf{J}_\mathrm{r}(\mathbf{r})$ on the helicoid and scattering current density $\mathbf{J}_\mathrm{s}(\mathbf{r})$ in the dielectric surrounding[4]:

$$\mathbf{J}_\mathrm{r}(\mathbf{r}) = -i\omega\varepsilon_0[\varepsilon_r - 1]\mathbf{E}(\mathbf{r}) \tag{S1}$$

$$\mathbf{J}_\mathrm{s}(\mathbf{r}) = -i\omega\varepsilon_0[\varepsilon_r - \varepsilon_h]\mathbf{E}(\mathbf{r}) \tag{S2}$$

First, the multipolar contribution in the macroscopic far-field response (*i.e.*, extinction) was decomposed by projecting the full induced current distribution onto the spherical multipole basis. The electric and magnetic spherical multipole coefficients, $a_E(l,m)$ and $a_M(l,m)$, were evaluated from $\mathbf{J}_\mathrm{s}(\mathbf{r})$ as:

$$a_E(l,m) = \frac{(-i)^{l-1}k^2\eta O_{lm}}{E_0\sqrt{\pi(2l+1)}}\int e^{-im\phi}\left\{[\Psi_l(kr) + \Psi_l''(kr)]P_l^m(\cos\theta)\,\hat{\mathbf{r}}\cdot\mathbf{J}_s + \frac{\Psi_l'(kr)}{kr}\left[\tau_{lm}(\theta)\,\hat{\boldsymbol{\theta}}\cdot\mathbf{J}_\mathrm{s} - i\pi_{lm}(\theta)\,\hat{\boldsymbol{\phi}}\cdot\mathbf{J}_\mathrm{s}\right]\right\}d^3r \tag{S3}$$

$$a_M(l,m) = \frac{(-i)^{l+1}k^2\eta O_{lm}}{E_0\sqrt{\pi(2l+1)}\int e^{-im\phi}j_l(kr)\left[i\pi_{lm}(\theta)\,\hat{\boldsymbol{\theta}}\cdot\mathbf{J}_\mathrm{s} + \tau_{lm}(\theta)\,\hat{\boldsymbol{\phi}}\cdot\mathbf{J}_\mathrm{s}\right]d^3r} \tag{S4}$$

where $k$ and $\eta$ are the wavenumber and impedance of the host medium, $j_l$ is the spherical Bessel function, and $\Psi_l(kr) = kr\,j_l(kr)$ is the Riccati-Bessel function. $P_l^m$, $\tau_{lm}$, $\pi_{lm}$, and $O_{lm}$ are the angular functions and normalization factor in the vector-spherical-harmonic representation. Multipole order is then determined by $l$, where $l = 1$ corresponds to the electric dipole (ED) and magnetic dipole (MD) modes and $l = 2$ corresponds to the electric quadrupole (EQ) and magnetic quadrupole (MQ) modes. Their extinction cross-section then can be evaluated from the modal overlap between the incident field and the induced multipole response,

$$C_\mathrm{ext} = -\frac{\pi}{k^2}\,\mathrm{Re}\left[\sum_{l,m}(2l+1)\big(b_E(l,m)^*a_E(l,m) + b_M(l,m)^*a_M(l,m)\big)\right] \tag{S5}$$

where $b_E(l,m)$ and $b_M(l,m)$are the incident-field expansion coefficients in the same spherical multipole basis. The result of calculation is displayed in Figure 3a, i.

We note that we interpret the origin of the chiroptical response in our system as being predominantly governed by the interference between the ED-MD throughout this work[5,6]. From a parity perspective, the ED is a polar vector with odd parity, whereas the MD is an axial vector with even parity. Their interference forms a pseudoscalar invariant that survives orientational averaging in colloidal ensembles, thereby enabling the differentiation in the interaction with light helicity. In contrast, while the electric quadrupole (EQ) also possesses even parity as MD, its interference with the ED undergoes a sign reversal under opposing orientations; consequently, these contributions cancel out at the ensemble level and do not yield a net macroscopic chiroptical signal. Accordingly, both the main text and subsequent discussions (Supplementary Notes 5 and 6) focus exclusively on the ED-MD modal interference.

Next, the multipole moment densities within the particle volume were expressed in terms of $\mathbf{J}_r(\mathbf{r})$ to visualize the local microscopic multipolar interference picture. The local ED and MD moment densities can be written as:

$$\mathbf{p}(\mathbf{r}) = \frac{i}{\omega}\mathbf{J}(\mathbf{r}) \tag{S6}$$

$$\mathbf{m}(\mathbf{r}) = \frac{1}{2}\,\mathbf{r} \times \mathbf{J}(\mathbf{r}) \tag{S7}$$

Numerical calculations confirm that the intricate geometry of the helicoid leads to a spatially varying mixing profile (*i.e.*, included angle) between the $\mathbf{p}(\mathbf{r})$ and $\mathbf{m}(\mathbf{r})$ vectors (Figure S9). Given that the inner product of these two components dictates the aspect of chiroptical response (see Supplementary Note 6), this spatial differentiation in their mixing underlies the localized handedness preference observed in field confinement. Specifically, such variations correlate with the circular contrast in the field distributions at the chiral gap and the periphery (Figure 3d), identifying this modal mixing as the microscopic origin of the system's spatially dependent chiroptical activity.

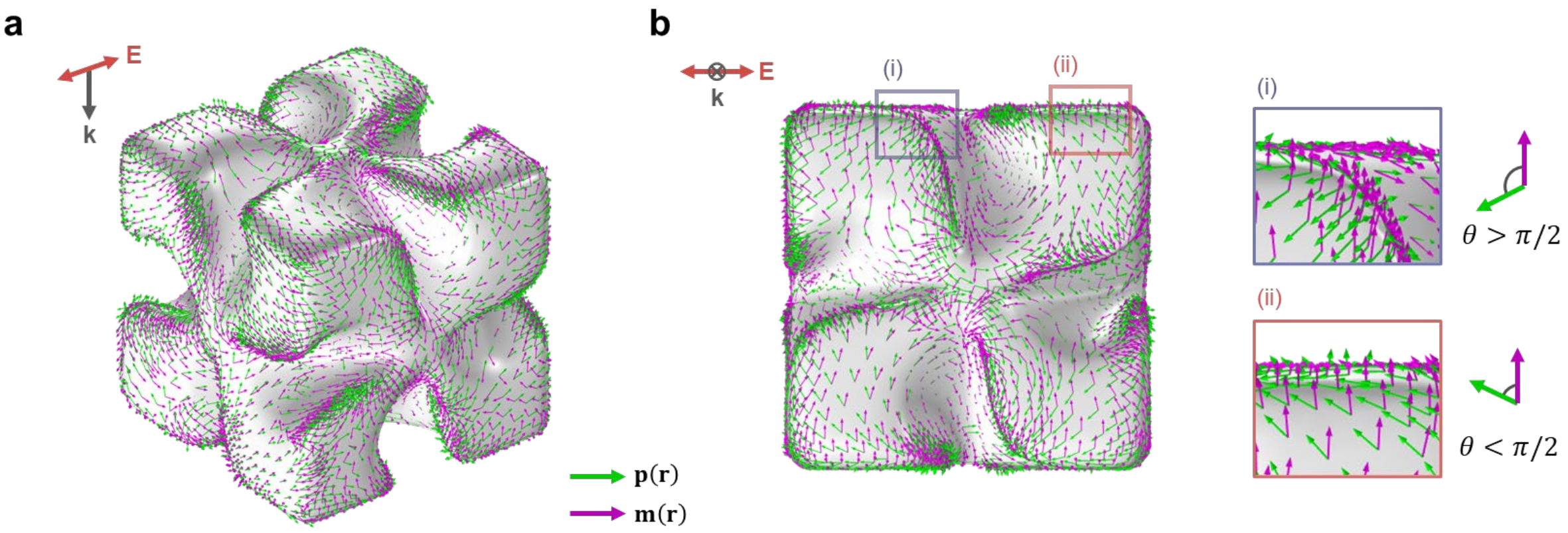

**Figure S9. Numerically simulated ED and MD moment densities.** (**a**) Perspective and (**b**) top views of the ED (green) and MD (magenta) moment densities on the plasmonic helicoid's surface. The distinct aspect of included angle $\theta$ between the ED and MD moment densities within the zoomed-in blue (i) and red (ii) boxes in (**b**) are correlated with the preferential addressment of RCP (i) and LCP (ii) light in those respective regions.

**Supplementary Note 5. NHH modelling for chiroptical response of plasmonic and plexcitonic helicoids**

To capture the pronounced plasmonic chirality driven by the ED-MD interplay and to quantify the fundamental nature of the coupling, we develop a theoretical framework for such interference based on non-Hermitian Hamiltonian (NHH). The NHH describing the plasmonic excitation in helicoid can be written as:

$$\begin{pmatrix} E_E - i\gamma_E & V_{EM}e^{i\Phi_{EM}} \\ V_{EM}e^{-i\Phi_{EM}} & E_M - i\gamma_M \end{pmatrix} \tag{S8}$$

where $E_E, E_M$ are the resonance energy of the 'bare' modes $|ED\rangle$ and $|MD\rangle$. $\gamma_{ED}, \gamma_{MD}$ are the broadening of the transitions, which describe the damping of the plasmon oscillations. $V_{EM}e^{\pm i\Phi_{EM}}$ is near-field electronic coupling the two modes. While the scalar $V_{EM}$ allows for the coherent hybridization of the plasmonic modes, $\Phi_{EM}$ is the relative phase between the ED and MD modes. In Born-Kuhn model of chiral plasmonic coupled oscillators, the phase retardance term depends on the spacing between coupled oscillators along the direction of light[7]. Here the relative phase $\Phi_{EM}$ represents the relative phase between the $ED$ and $MD$ plasmonic oscillations in the chiral spatial geometry of the helicoid. The two terms $V_{EM}, \Phi_{EM}$ along with the relative orientation of the TDMs determine the strength and sign of optical rotational strength in the helicoid (see Supplementary Note 6).

The eigenvalues λ are:

$$\varepsilon_k = \left(\frac{E_E + E_M}{2}\right) - i\left(\frac{\gamma_E + \gamma_M}{2}\right) \pm \frac{1}{2}\sqrt{4V_{EM}^2 - \left(\frac{\Delta E - i\Delta\gamma}{2}\right)^2} \tag{S9}$$

Where $\Delta E = E_E - E_M$ and $\Delta\gamma = \gamma_E - \gamma_M$. $k = 1,2$

The eigenstates $|\psi_k\rangle$ are given by $|\psi_k\rangle = a_k|ED\rangle + b_k|MD\rangle$ Where $a_k, b_k$ are complex scalars and $k = 1,2$. To calculate the far-field optical absorption, we require the TDMs of the system $\mu_k$ and $m_k$:

$$\mu_k = \langle 0|\hat{\mu}|\psi_k\rangle = a_k\mu_{ED} + b_k\mu_{MD} \tag{S10}$$

$$m_k = \langle 0|\widehat{m}|\psi_k\rangle = a_k m_{ED} + b_k m_{MD} \tag{S11}$$

$\mu_{ED} = \langle 0|\hat{\mu}|ED\rangle$, $\mu_{MD} = \langle 0|\hat{\mu}|MD\rangle$ are the electric TDMs of the bare modes; $m_{ED} = \langle 0|\widehat{m}|ED\rangle, m_{MD} = \langle 0|\widehat{m}|MD\rangle$ are the magnetic TDMs of the bare modes. We can argue that the TDMs $\mu_{MD}$ and $m_{ED}$ are zero.

The RCP and LCP optical response is proportional to $(\mu + im)^2$ and $(\mu - im)^2$.[8] We assume that the attenuation of light or linear optical response can be approximated by Lorentzian line shapes, as was done in previous work[9,10]. The optical attenuation of LCP ($\alpha_+$) and RCP ($\alpha_-$) is written as:

$$\alpha_\pm(E) = c + \sum_{k=1}^{2} \frac{[\mathrm{Re}((\mu_k \pm im_k)^2) * \mathrm{Im}(\varepsilon_k)] - [\mathrm{Im}((\mu_k \pm im_k)^2) * (E - \mathrm{Re}(\varepsilon_k))]}{[E - \mathrm{Re}(\varepsilon_k)]^2 + [\mathrm{Im}(\varepsilon_k)]^2} \tag{S12}$$

$c$ is a constant that offsets the background from signal. The attenuation of linear polarized light is the average of the two, given as $(\alpha_{+}+\alpha_{-})/2$.

The fitted parameters for the helicoids are given in Table S1. The obtained coupling terms are $V_{EM} =$ 46 meV for $Helicoid_{(A)}$ and $V_{EM} = 31$ meV for $Helicoid_{(B)}$.

**Table S1. Plasmon coupling model of helicoid**

| Parameter | Helicoid(A) | Helicoid(B) | Description |
|---|---|---|---|
| $E_E$ | $2.141 \pm 0.001\ eV$ | $2.094 \pm 0.001\ eV$ | ED plasmon mode energy |
| $E_M$ | $1.914 \pm 0.002\ eV$ | $1.801 \pm 0.001\ eV$ | MD plasmon mode energy |
| $\gamma_E$ | $0.275 \pm 0.003\ eV$ | $0.348 \pm 0.003\ eV$ | ED plasmon mode broadening |
| $\gamma_M$ | $0.126 \pm 0.002\ eV$ | $0.152 \pm 0.002\ eV$ | MD plasmon mode broadening |
| $V_{EM}$ | $0.046 \pm 0.001\ eV$ | $0.030 \pm 0.001\ eV$ | ED-MD mode coupling |
| $\mu_{ED}$ | $0.185 \pm 0.002$ | $0.225 \pm 0.002$ | ctric transition dipole moment bare ED plasmon |
| $m_{MD}$ | $0.049 \pm 0.001$ | $0.070 \pm 0.001$ | gnetic transition dipole moment MD plasmon mode |
| $\Phi_{EM}$ | $(-0.983 \pm 0.005)\pi\ rad$ | $(0.979 \pm 0.003)\pi\ rad$ | iral geometric phase between E and MD modes |

The deviation of geometric phase $\Phi_{EM}$ from $n\pi$ reflects the breaking of orthogonality of electric and magnetic dipole transitions, leading to the circular dichroism. In the case of helicoid, the geometry forces the breaking of this orthogonality.

In order to obtain the plexcitonic lineshapes we perform a straightforward extension to a 3x3 NHH. In the plexciton, we have the already established hybrid plasmon system (helicoid) interacting with an excitonic (TDBC J-aggregate) system. The Hamiltonian can be written as:

$$\begin{pmatrix} E_E - i\gamma_E & V_{EM}e^{i\Phi_{EM}} & V_{EJ}e^{i\Phi_{EJ}} \\ V_{EM}e^{-i\Phi_{EM}} & E_M - i\gamma_M & 0 \\ V_{EJ}e^{-i\Phi_{EJ}} & 0 & E_J - i\gamma_J \end{pmatrix} \tag{S13}$$

Here, we define the new coupling term, $V_{EJ}$ as the coupling strength of the bare plasmonic mode $|ED\rangle$ to the J-aggregate $|J\rangle$ and $\Phi_{EJ}$ is the geometric phase between the J-aggregate and the plasmonic mode $|ED\rangle$ in the near-field. It is assumed that there is no coupling between the states $|MD\rangle$ and $|J\rangle$. The eigenstates are now obtained as $|\Psi_k\rangle = a_k|ED\rangle + b_k|MD\rangle + c_k|J\rangle$. The TDMs for the uncoupled J-aggregate are written as $\mu_J = \langle 0|\hat{\mu}|J\rangle$ and $m_J = \langle 0|\hat{m}|J\rangle$ and for the TDMs for the coupled system:

$$\mu_k = a_k\mu_{ED} + c_k\mu_J \tag{S14}$$

$$m_k = b_k m_{MD} + c_k m_J \tag{S15}$$

As discussed after equations (S10) and (S11), we have neglected the induced dipole moments $\mu_{MD}$ and $m_{ED}$. $\mu_J$ is included to account for the electric TDM expected for the J-aggregate absorption.

The results of the fitting are given in Table S2 and the obtained fits are shown in Figure 3c. The term describing plasmon-exciton coupling is $V_{EJ} = 42$ meV for Helicoid$_{(A)}$ and $V_{EJ} = 45$ meV for Helicoid$_{(B)}$. whereas $V_{MJ} = 0$. As described in the main text, the extracted coupling strength $(g)$ places the Fano-like interfering mode in an intermediate coupling regime. While insufficient for strong coupling, it exceeds weak coupling limits, driving the observed far-field spectral splitting (Figure S12).

**Table S2. Plexcitonic Coupling Model of Helicoid-Dye**

| Parameter | Helicoid(A)-TDBC | Helicoid(B)-TDBC | Description |
|---|---|---|---|
| $E_E$ | $2.141 \pm 0.001\ eV$ | $2.094 \pm 0.001\ eV$ | ED plasmon mode energy |
| $E_M$ | $1.866 \pm 0.002\ eV$ | $1.776 \pm 0.001\ eV$ | MD plasmon mode energy |
| $E_J$ | $2.092 \pm 0.001\ eV$ | $2.092 \pm 0.001\ eV$ | J-aggregate exciton energy |
| $\gamma_E$ | $0.275 \pm 0.003\ eV$ | $0.405 \pm 0.003\ eV$ | ED plasmon mode broadening |
| $\gamma_M$ | $0.126 \pm 0.002\ eV$ | $0.135 \pm 0.002\ eV$ | MD plasmon mode broadening |
| $\gamma_J$ | $0.020 \pm 0.001\ eV$ | $0.020 \pm 0.001\ eV$ | Dye exciton broadening |
| $V_{EM}$ | $0.046 \pm 0.001\ eV$ | $0.030 \pm 0.001\ eV$ | ED-MD mode coupling |
| $V_{EJ}$ | $0.042 \pm 0.001\ eV$ | $0.045 \pm 0.001\ eV$ | ED-J-aggregate coupling |
| $V_{MJ}$ | 0 | 0 | MD-J-aggregate coupling |
| $\mu_{ED}$ | $0.185 \pm 0.003$ | $0.269 \pm 0.002$ | tric transition dipole moment of b ED plasmon |
| $m_{MD}$ | $0.049 \pm 0.001$ | $0.070 \pm 0.001$ | ıgnetic transition dipole moment bare MD plasmon |
| $\mu_J$ | $0.010 \pm 0.001$ | $0.010 \pm 0.001$ | ctric transition dipole moment of aggregate |
| $m_J$ | $0.004 \pm 0.001$ | $0.001 \pm 0.001$ | gnetic transition dipole moment o aggregate |
| $\Phi_{EM}$ | $(-0.982 \pm 0.002)\pi\ rad$ | $(0.953 \pm 0.002)\pi\ rad$ | al geometric phase between ED MD modes |
| $\Phi_{EJ}$ | $(-0.071 \pm 0.002)\pi\ rad$ | $(-0.114 \pm 0.002)\pi\ rad$ | metric phase between ED near-fi and J-aggregate |

The chiroptical response of the plexcitonic system is determined by the two coupling terms $V_{EM}, V_{EJ}$ and the geometric phases $\Phi_{EM}, \Phi_{EJ}$. While the theoretical model was successful in reproducing the optical response of both bare helicoid and helicoid-TDBC aggregates, two points were noted while interpreting the model result:

1. The fit yielded a strong correlation between the obtained parameters $\Phi_{EJ}, V_{EJ}$ and $\mu_J$, which indicates that the actual uncertainty in these values was higher, since they are clearly inter-dependent. However, there exists a considerable ED plasmon-J-aggregate coupling. This is also reflected in the differential time-domain relaxation dynamics of the plexciton vs plasmon in Figure 4 in Main text.

2. It is to be noted that the J-aggregate having a magnetic TDM $m_J \neq 0$ is not expected physically. Yet it is found that including a small non-zero magnetic TDM $m_J$ yields a good fit of the linear response near the J-aggregate energy. The non-zero value of $m_J$ is interpreted as arising due to a possible enhancement of the J-aggregate-ED plasmon coupling $V_{EJ}$ at the J-aggregate resonance energy. The model considers $V_{EJ}$ as independent of frequency, which might not be the case at the J-aggregate resonance energy.

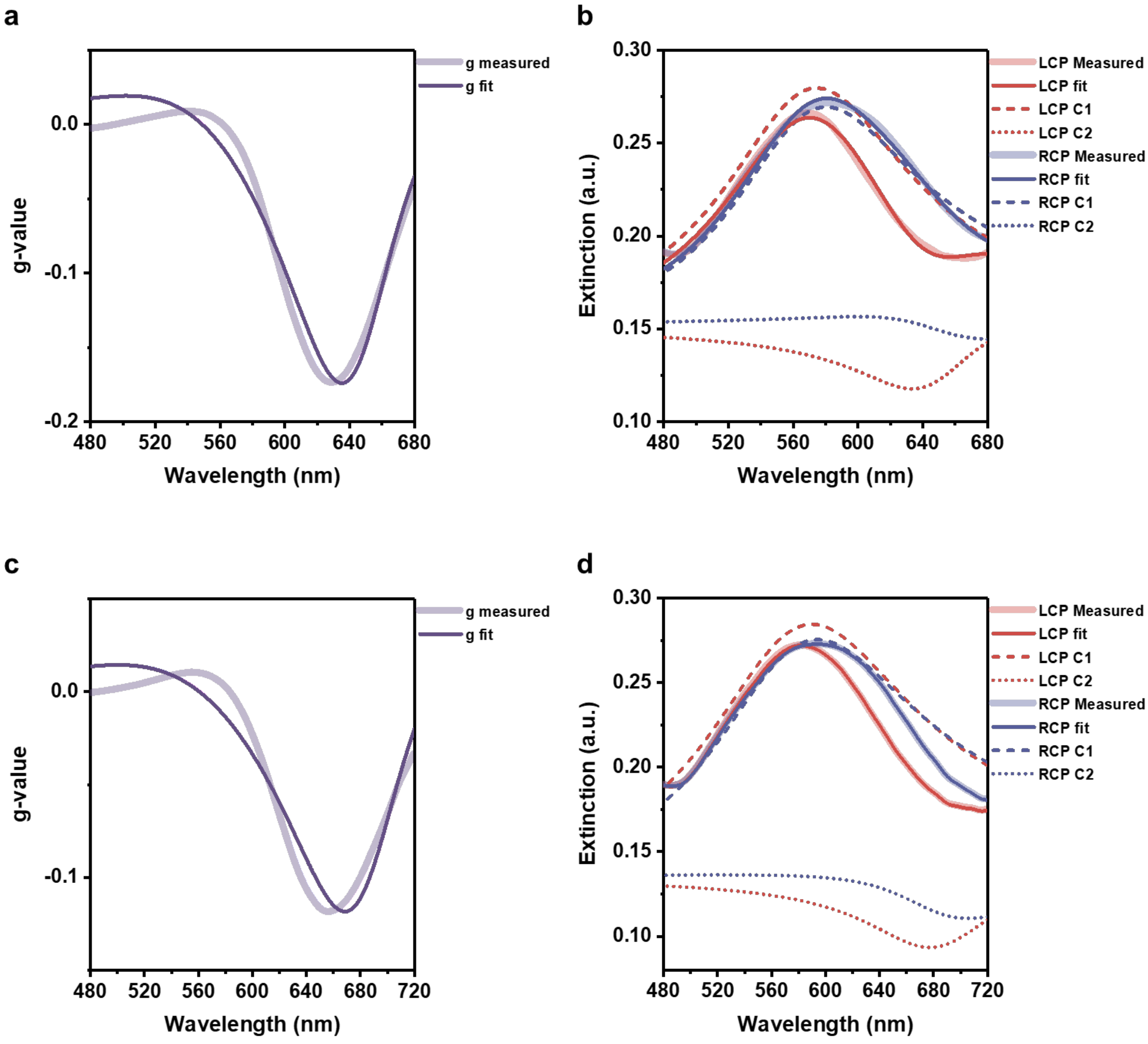

**Figure S10. NHH model fit of plasmonic helicoids. a,** *g*-factor and **b,** extinction spectra of pristine Helicoid$_{(A)}$. **c,** *g*-factor and **d,** extinction spectra of pristine Helicoid$_{(B)}$. Experimental data (blurred lines) and overall fits (solid lines) are shown in both panels, with individual modal contributions (broken lines) additionally plotted in (**b** and **d**).

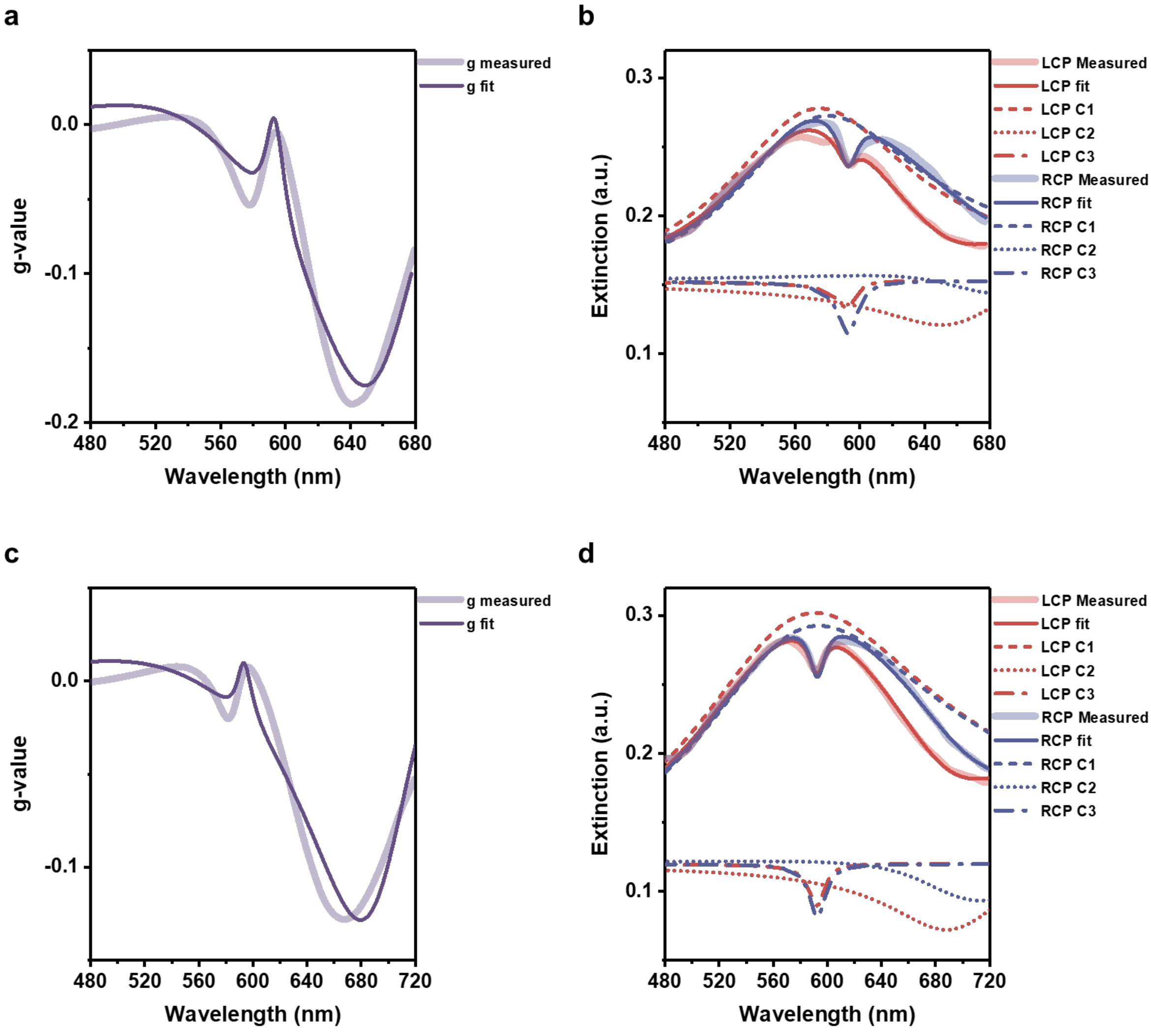

**Figure S11. NHH model fit of plexcitonic helicoids. a,** *g*-factor and **b,** extinction spectra of Helicoid$_{(A)}$-TDBC conjugate. **c,** *g*-factor and **d,** extinction spectra of Helicoid$_{(B)}$-TDBC conjugate. Experimental data (blurred lines) and overall fits (solid lines) are shown in both panels, with individual modal contributions (broken lines) additionally plotted in (**b** and **d**).

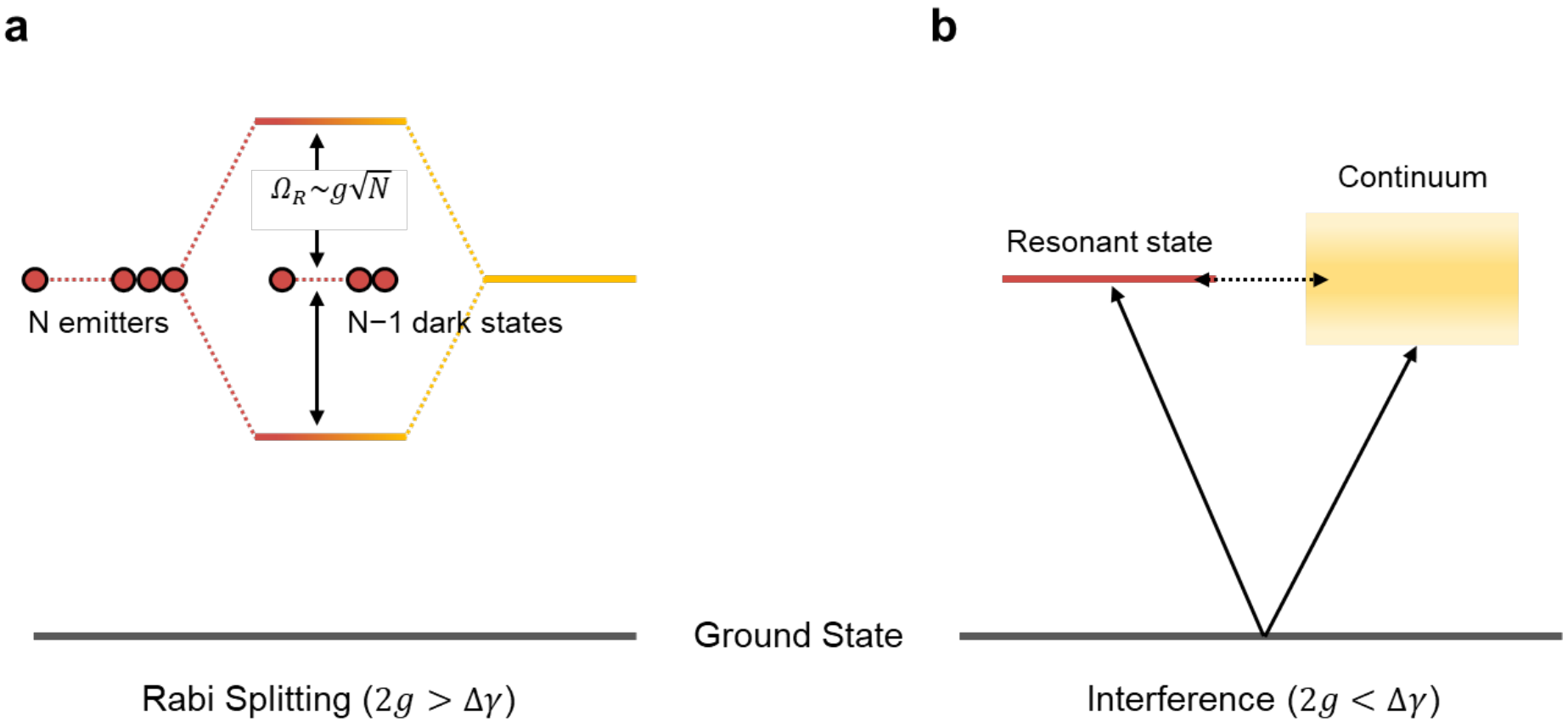

**Figure S12. Schematic illustration representing plasmon-exciton coupling regime. a,** Strong coupling between an ensemble of N two-level systems (red) with an electromagnetic field (yellow) in (Tavis-Cummings model). Rabi-split eigenstates are formed, with the magnitude of splitting = $g\sqrt{N}$ where $g$ is the light-matter coupling strength. **b,** Intermediate coupling that is relevant to plexcitons in this work, between a lossy plasmon mode and a resonant exciton state creates interference in the optical response, leading to Fano lineshapes.

## Supplementary Note 6. Optical rotational strength of helicoids

The optical rotational strength ($R_n$, for a transition from $|0\rangle$ to $|n\rangle$) is described by the Rosenfeld relation:

$$R_n = \mathrm{Im}(\mu_{0n}.m_{0n}^*) = \mathrm{Im}(\langle 0|\mu|n\rangle.\langle n|m|0\rangle) \tag{S16}$$

Where $\mu$ and $m$ are the electric and magnetic TDMs. In the helicoids the modes $|ED\rangle$ and $|MD\rangle$ are coupled via the term $V_{EM}e^{i\Phi_{EM}}$. The coupling gives rise to mixed eigenstates that are expressed as $|\psi_k\rangle = a_k|ED\rangle + b_k|MD\rangle$ Where $a_k, b_k$ are complex scalars and $k = 1,2$. The rotational strength for transition to the k$^{th}$ state is then:

$$R_k = \mathrm{Im}(\langle 0|\mu|\psi_k\rangle\langle\psi_k|m|0\rangle) = \mathrm{Im}(a_k\mu_{ED}.b_k^* m_{MD}) \tag{S17}$$

$$R_k = \mu_{ED}.m_{MD}\,\mathrm{Im}(a_k b_k^*) \tag{S18}$$

For the helicoid to have non-zero rotational strength:[1] 1. The orientations of $\mu_{ED}$ and $m_{MD}$ need to be non-orthogonal, and 2. $\mathrm{Im}(a_k b_k^*) \neq 0$. To evaluate $\mathrm{Im}(a_k b_k^*)$, we start from the Hamiltonian (equation (S8)). Complex energy of the modes are defined as: $\widetilde{E_E} = E_E - i\gamma_E$, $\widetilde{E_M} = E_M - i\gamma_M$ and their detuning $P = \widetilde{E_E} - \widetilde{E_M} = (E_E - E_M) - i(\gamma_E - \gamma_M)$.

The eigenstates are $|\psi_1\rangle, |\psi_2\rangle$ with complex eigenvalues $\widetilde{E_1}, \widetilde{E_2}$. The eigenvalue equation is $(H - \widetilde{E_1})|\psi_1\rangle = 0$. Solving the second row:

$$V_{EM}e^{-i\Phi_{EM}}a_1 + (\widetilde{E_M} - \widetilde{E_1})b_1 = 0 \tag{S19}$$

A perturbative treatment of the eigenstates for small $V_{EM} < |\widetilde{E_M} - \widetilde{E_E}|$, which is a reasonable approximation in the context of helicoids (Values in Table S1), implies that $\widetilde{E_1} \approx \widetilde{E_E}$:

$$V_{EM}e^{-i\Phi_{EM}}a_1 + (\widetilde{E_M} - \widetilde{E_E})b_1 = 0 \tag{S20}$$

$$b_1 = \left(\frac{V_{EM}e^{-i\Phi_{EM}}}{(\widetilde{E_E} - \widetilde{E_M})}\right)a_1 \tag{S21}$$

For $V_{EM} < |\widetilde{E_M} - \widetilde{E_E}|$,

$$a_1 \approx 1\,;\; b_1 \approx \left(\frac{V_{EM}e^{-i\Phi_{EM}}}{(\widetilde{E_E} - \widetilde{E_M})}\right) \tag{S22}$$

$$\mathrm{Im}(a_1 b_1^*) = \mathrm{Im}\left(\frac{V_{EM}e^{+i\Phi_{EM}}}{(\widetilde{E_E} - \widetilde{E_M})^*}\right) = \mathrm{Im}\left(\frac{V_{EM}(\cos(\Phi_{EM}) + i\sin(\Phi_{EM}))}{(E_E - E_M) + i(\gamma_E - \gamma_M)}\right) \tag{S23}$$

$$\mathrm{Im}(a_1 b_1^*) = \mathrm{Im}\left(\frac{V_{EM}(\cos(\Phi_{EM}) + i\sin(\Phi_{EM}))\big((E_E - E_M) - i(\gamma_E - \gamma_M)\big)}{(E_E - E_M)^2 + (\gamma_E - \gamma_M)^2}\right) \tag{S24}$$

$$\mathrm{Im}(a_1 b_1^*) = V_{EM}\left(\frac{(E_E - E_M)\sin(\Phi_{EM}) - (\gamma_E - \gamma_M)\cos\Phi_{EM}}{(E_E - E_M)^2 + (\gamma_E - \gamma_M)^2}\right) \tag{S25}$$

Let $(E_E - E_M) = \Delta E$ and $(\gamma_E - \gamma_M) = \Delta\gamma$ and $\delta = \tan^{-1}\left(\frac{\Delta\gamma}{\Delta E}\right)$:

$$\mathrm{Im}(a_1 b_1^*) = V_{EM}[\sin(\Phi_{EM})\cos(\delta) + \cos(\Phi_{EM})\sin(\delta)] \tag{S26}$$

$$\mathrm{Im}(a_1 b_1^*) = V_{EM}[\sin(\Phi_{EM} + \delta)] \tag{S27}$$

So the rotational strength of the transition to mode 1 is:

$$R_1 = (\mu_{ED}.m_{MD})V_{EM}\,[\sin(\Phi_{EM} + \delta)] \tag{S28}$$

The rotational strength is proportional to the $g$. The magnitude of the rotational strength depends on the dot product $(\mu_{ED}.m_{MD})$ and the coupling $V_{EM}$. The sign of rotational strength depends on the geometric phase and the detuning factor $\delta$, with $\Phi_{EM} < n\pi - \delta$ giving negative sign and $\Phi_{EM} > n\pi - \delta$ giving a positive sign.

## Supplementary Note 7. Modelling and numerical implementation of TDBC optical constants

To numerically simulate the optical response of the helicoid-TDBC system, the excitonic optical properties of the TDBC J-aggregates were analytically modelled and subsequently implemented into our computational framework. Dispersion of optical constant $\varepsilon_\omega$ of TDBC J-aggregate was modeled by Lorentz oscillator model:

$$\varepsilon_\omega = \varepsilon_\infty - \frac{f\omega_0^2}{\omega^2 - \omega_0^2 + i\Gamma\omega} \tag{S29}$$

where oscillator strength f was set to be 0.015 and all other parameters consisting of the equation were taken referring to the literature[11]. The dispersion of $\varepsilon_\omega$ derived in this manner accurately captures the behavior of the excitonic band at 590 nm (Figure S13a).

The structure wrapping this on the gold helicoid as a core-shell structure exhibited a distinct peak split in the extinction cross section obtained through simulation. Under the identical simulation setup as in Supplementary Note 4, we applied a 5-nm-thick conformal coating of the excitonic layer on the helicoid. The extinction cross-section was obtained as the sum of the absorption cross-sections $C_{abs}$ and scattering cross-sections $C_{sca}$[12]. The absorption cross-section was evaluated from the volume-integrated resistive losses within the particle:

$$C_{abs} = \frac{\int_V \frac{1}{2}\mathrm{Re}[\mathbf{J}_\mathrm{r}(\mathbf{r}) \cdot \mathbf{E}^*(\mathbf{r})]dV}{I_{inc}} \tag{S30}$$

while the scattering cross-section was determined by integrating the scattered Poynting vector over the boundary of the computational domain:

$$C_{sca} = \frac{\oint_{\partial\Omega} \left\{\frac{1}{2}\mathrm{Re}[\mathbf{E}_{sca}(\mathbf{r}) \times \mathbf{H}_{sca}^*(\mathbf{r})]\right\} \cdot \hat{\mathbf{n}} dA}{I_{inc}} \tag{S31}$$

The splitting spectral feature in the extinction cross-section spectrum verifies that the coupling induced optical behaviour was successfully captured (Figure S13b).

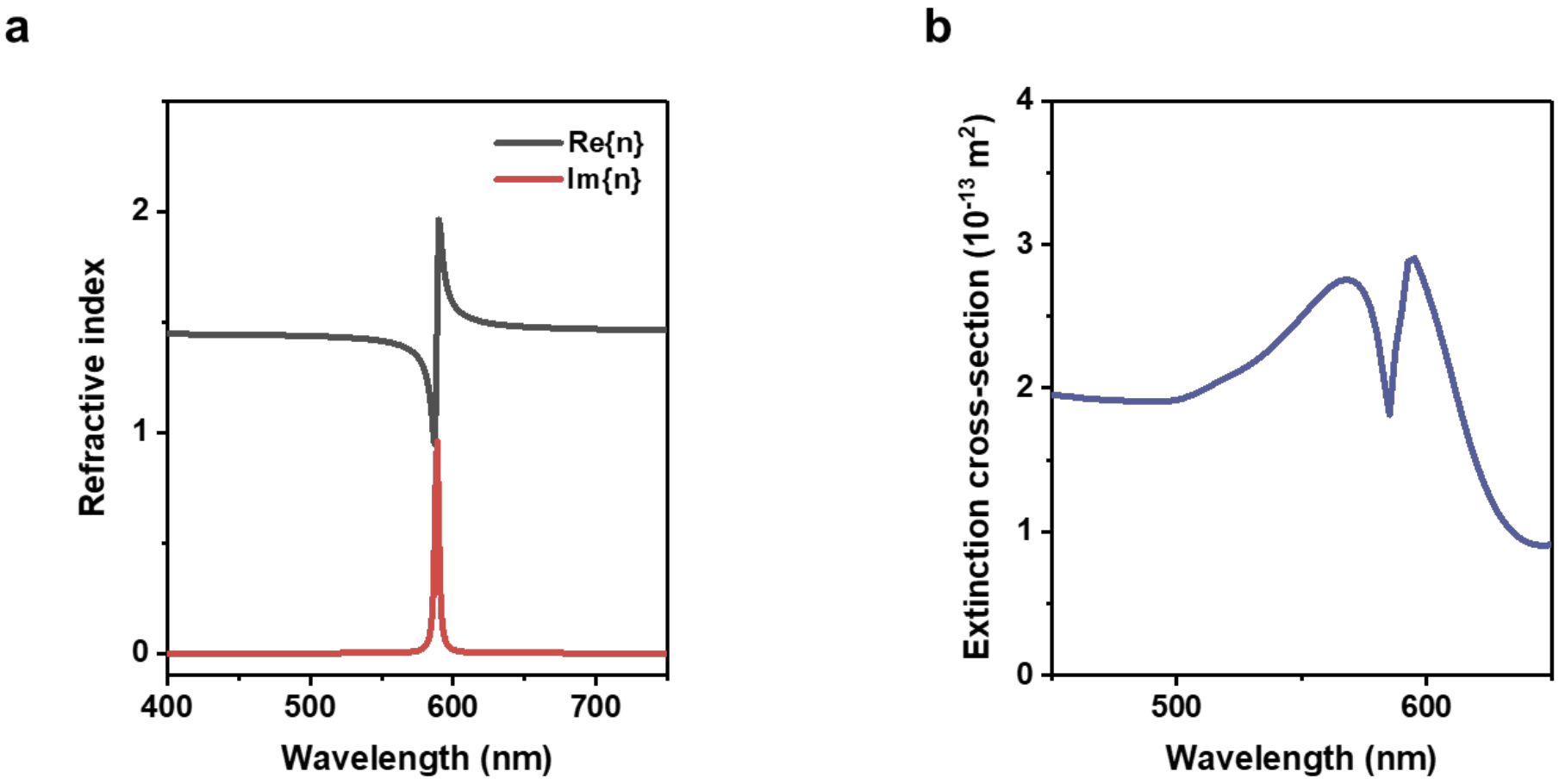

**Figure S13. Modelling and numerical implementation of TDBC optical constants. a,** Real and imaginary parts of the refractive index of TDBC, derived from the Lorentz oscillator model. **b,** Numerically simulated extinction cross-section of helicoid-TDBC core-shell structure.

## Supplementary Note 8. Probe-pulse polarization controlled-TA measurements

The pseudo-colour TA maps for both the pristine and TDBC-conjugated helicoids, measured across varying probe polarizations, are displayed in Figures S14 and S15. As evidenced by steady-state optical measurements, the incident polarization modulates both the spectral peak position and the linewidth of the helicoid's plasmonic band. This polarization dependence directly translates into the pump-probe regime, with these distinct linewidth characteristics (RCP > LP > LCP) are prominently manifested in the TA maps of both the bare plasmonic and plexcitonic systems. Spectral slices extracted from the LP-probed TA maps (Figures S14b and S15b) are plotted in Figures 4b and 4c.

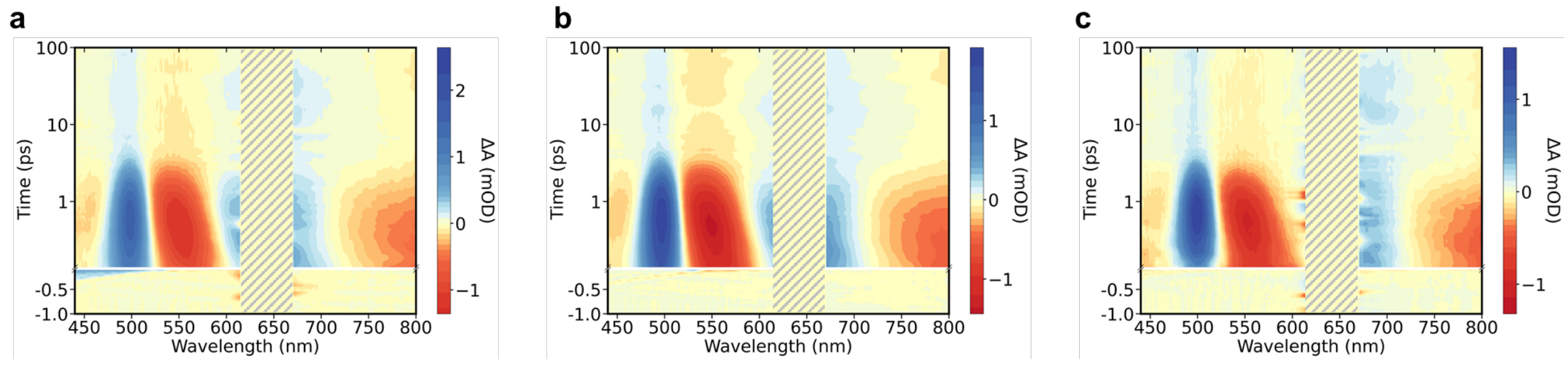

**Figure S14. Pseudo-colour TA maps of pristine helicoid acquired at varying pump powers and probe polarizations. a-c,** TA maps obtained under probe polarizations of LCP (**a**), LP (**b**), and RCP (**c**). Hatched region corresponds to the spectral area obscured by pump scattering.

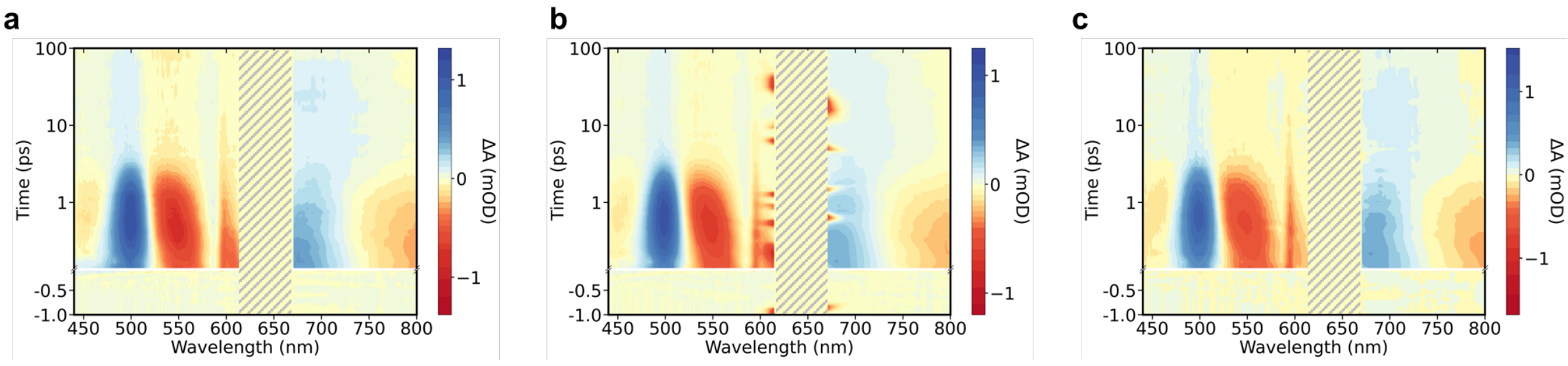

**Figure S15. Pseudo-colour TA maps of helicoid-TDBC conjugate acquired at varying probe polarizations.** **a-c,** TA maps obtained under probe polarizations of LCP (**a**), LP (**b**), and RCP (**c**). Hatched region corresponds to the spectral area obscured by pump scattering.

**Supplementary Note 9. Mechanistic study on the plexcitonic relaxation pathways via TA measurements**

For a mechanistic investigation into the energy redistribution pathways induced by plexcitonic coupling, we compared the photobleaching dynamics in the two systems (*i.e.*, pristine and TDBC-conjugated helicoids) by acquiring TA spectra across a range of pump powers. Figure S16 shows the normalised dynamics of the plasmon at 550 nm, plexciton at 550 nm (UB) and 590 nm (LB), respectively, at 1 mW, 4 mW, 8.5 mW and 13.3 mW of pump laser power. The dynamics are fitted with the equation based on a two-temperature model (TTM)[9,13]:

$$\Delta A = A\left(1 - \exp\left(-\frac{t - t_0}{\tau_{ee}}\right)\right) * \left(\exp\left(-\frac{t - t_0}{\tau_{ep}}\right)\right) \tag{S32}$$

where A is the amplitude, $\tau_{ee}$ is the thermalisation time of the electronic distribution and $\tau_{ep}$ is the electron-phonon scattering time, which determines the rate of energy loss from free electrons to the lattice. The result of the fitting is given in Table S3-5.

Two key observations from this analysis are as follows: (1) As excitation power increases, the electron thermalisation rate gets faster while the electron-phonon scattering rate is slower. The former can be explained by a higher rate of electron-electron scattering at a high density of free electrons in the system. The latter can be rationalised as due to a hot phonon bottleneck effect in the system, where higher excitation leads to a large phonon population that reheats the electron distribution. (2) Over the range of excitation powers, a faster rate of electron thermalisation is observed for the LB state compared to the UB state and the plasmon. The optical response at the lower plexciton energy originates mainly from the states located at the gaps, as presented in Figure 3e and f. Drawing upon previously reported relaxation pathways in plexcitonic systems, we propose that the accelerated relaxation of the LB state stems from the interaction of the plexciton with reservoir dark states (DS) near the gap. Specifically, as briefly illustrated in Figure S17, (i) non-thermalized electrons after plasmon damping transfer their energy by coupling to the DS of J-aggregates on a sub-picosecond timescale—a behaviour reminiscent of the known ultrafast relaxation of TDBC J-aggregate excitons into dark states. This shares the dynamic timescale with electron-electron scatterin. Then, (ii) over a longer timescale of tens of picoseconds, this DS reservoir acts as a thermal source, injecting energy back into the gold lattice until equilibrium is reached. The UB state exhibits a thermalization rate comparable to that of the bare plasmon, indicating a predominantly plasmonic character that remains largely unaffected by this dark-state-mediated pathway.

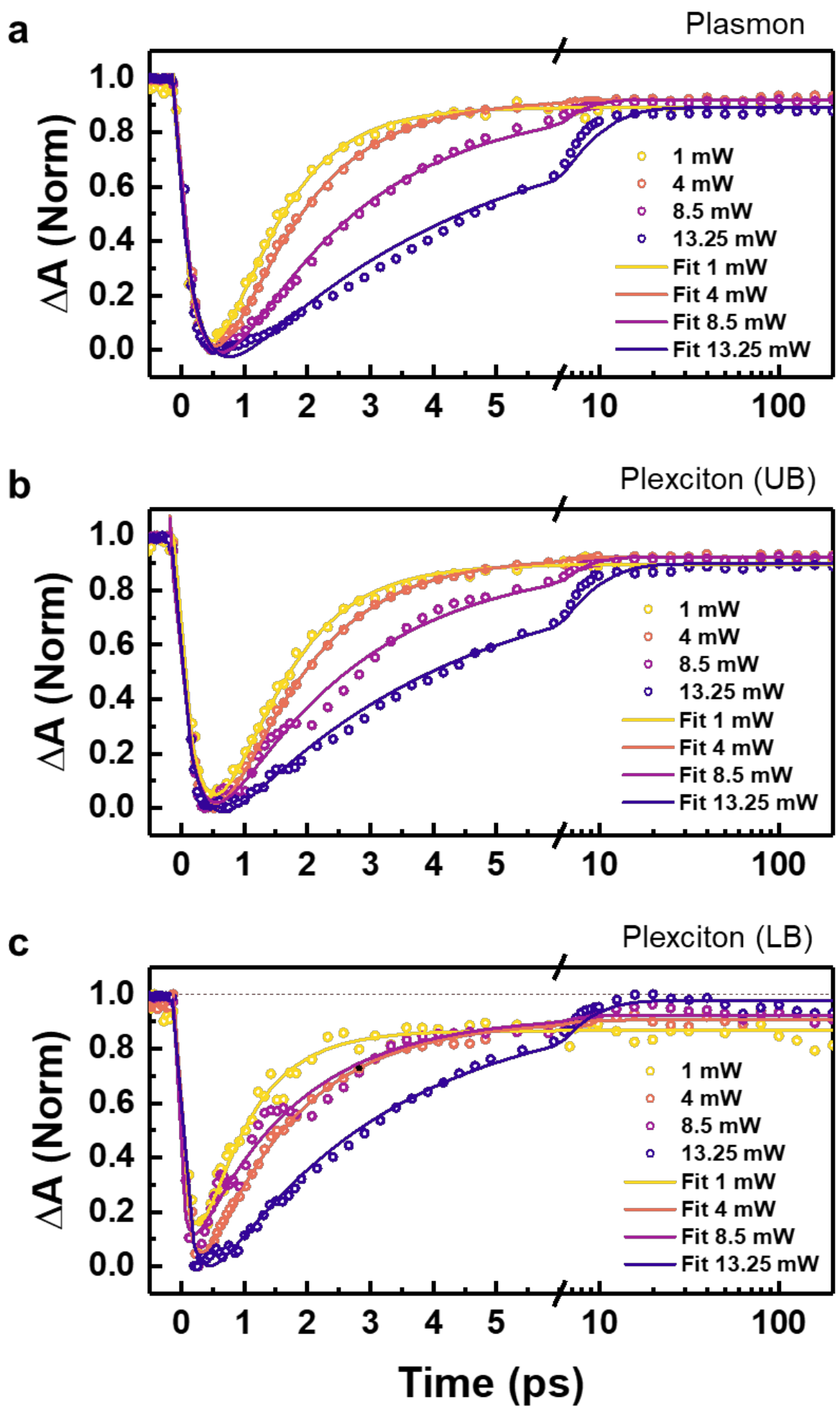

**Figure S16. Pump power dependence of TA dynamics.** Normalised power-dependent kinetics of **a,** plasmonic helicoid at 550 nm, **b,** helicoid-TDBC conjugate at 550 nm (UB), and **c,** helicoid-TDBC conjugate at 590 nm (LB).

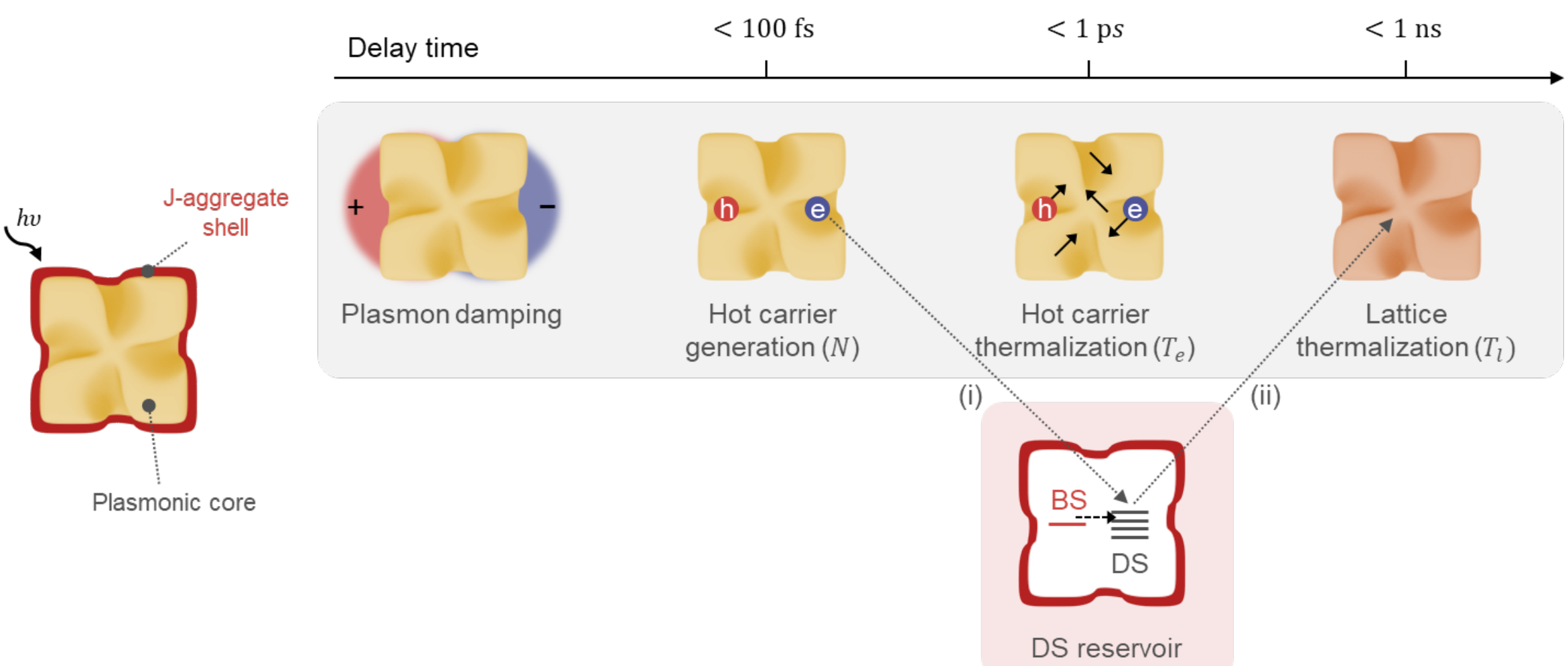

**Figure S17. Schematic illustration of ultrafast plexcitonic dynamics.** Following initial optical excitation, the plasmonic component undergoes fast damping to generate hot electrons (< 100 fs). These electrons thermalize ($T_e$) through electron-electron scattering (< 1 ps) and eventually dissipate energy to the metal lattice ($T_l$) via electron-phonon scattering (< 1 ns). Alongside this intrinsic pathway, plexcitonic coupling with J-aggregate introduces an additional dynamic pathway. (i) The excitonic dark states (DS) of J-aggregate act as a high-density reservoir, efficiently capturing hot electrons and accelerating their initial relaxation (< 1 ps). (ii) The temporarily stored energy is then back-transferred from the DS reservoir to the lattice, restoring the reservoir to equilibrium.

**Table S1: TTM fit for plasmon resonance in pristine helicoid**

| Plasmon | 1 mW | 4 mW | 8.5 mW | 13.3 mW |
|---|---|---|---|---|
| A (mO.D.) | -2.2 ± 0.4 | -6.7 ± 0.3 | -12.3 ± 0.2 | *-16.8 ± 0.3* |
| $\tau_{ee}$ *(ps)* | 0.82 ± 0.20 | 0.59 ± 0.04 | 0.40 ± 0.02 | 0.34 ± 0.02 |
| $\tau_{ep}$ *(ps)* | 0.85 ± 0.07 | 1.19 ± 0.04 | 2.18 ± 0.05 | 3.93 ± 0.18 |

**Table S2: TTM fit for upper branch (UB) in helicoid-TDBC conjugate**

| Plasmon | 1 mW | 4 mW | 8.5 mW | 13.3 mW |
|---|---|---|---|---|
| A (mO.D.) | -1.6± 0.4 | -4.2 ± 0.4 | -5.8± 0.2 | -8.7 ± 0.2 |
| $\tau_{ee}$ *(ps)* | 0.96 ± 0.27 | 0.63 ± 0.07 | 0.35 ± 0.03 | 0.30 ± 0.02 |
| $\tau_{ep}$ *(ps)* | 0.89 ± 0.08 | 1.23 ± 0.06 | 2.29 ± 0.11 | 3.64 ± 0.11 |

**Table S3: TTM fit for lower branch (LB) in helicoid-TDBC conjugate**

| Plasmon | 1 mW | 4 mW | 8.5 mW | 13.3 mW |
|---|---|---|---|---|
| A (mO.D.) | -0.4± 0.1 | -1.2 ± 0.1 | -3.0± 0.1 | -5.6 ± 0.1 |
| $\tau_{ee}$ *(ps)* | 0.27 ± 0.07 | 0.21 ± 0.02 | 0.12 ± 0.02 | 0.22 ± 0.02 |
| $\tau_{ep}$ *(ps)* | 0.78 ± 0.09 | 1.46 ± 0.05 | 1.64 ± 0.11 | 2.97 ± 0.10 |

## Supplementary Note 10. Numerical modelling on plasmonic and plexcitonic dynamics

Spatiotemporal dynamics were simulated based on the temporal dissipation dynamics of optical gain in the plasmon-based system. For numerical modelling, extended two temperature model (eTTM) on the typical plasmonic dynamics[14] and two partial differential equations (PDEs) for excitonic bright state (BS) and dark state (DS) containing the intercorrelation terms were simultaneously solved in the simulator by allotting the initial thermal energy retrieved from wave optics simulation. Optically injected time-averaged thermal energy density was retrieved in the simulation setup discussed in Supplementary Note 7, by solely considering the resistive loss density, neglecting magnetic loss density, based on following expression:

$$Q(\mathbf{r}) = \frac{1}{2}\mathrm{Re}[\mathbf{J}_{\mathrm{r}}(\mathbf{r}) \cdot \mathbf{E}^{*}(\mathbf{r})] \tag{S33}$$

We note that we averaged $Q(\mathbf{r})$ obtained under LCP and RCP illumination to eliminate the inherent anisotropy bias. To model the temporal injection profile of pump pulse, the magnitude was scaled considering our experimental pump fluence condition:

$$F = \frac{P}{\pi r^2 f_{rep}} \tag{S34}$$

where $P$ is time-averaged power (13.3 mW), $r$ is beam radius (200 μm), and $f_{rep}$ is repetition rate (2 kHz). Then it was again scaled in time domain assuming Gaussian pulse:

$$A(t) = \exp\left[-4\ln 2\left(\frac{t}{\Delta t}\right)\right] \tag{S35}$$

with $\Delta t = 100$ fs pulse width. Then, the spatiotemporal evolution of initial optical gain was calculated by solving the following PDEs:

$$\frac{dN}{dt} = \frac{1}{C_e(T_e)}\nabla \cdot [\kappa_e \nabla N] - (a + b + k_{N-DS})N + Q \cdot A(t) \tag{S36}$$

$$C_e(T_e)\frac{dT_e}{dt} = \nabla \cdot [\kappa_e \nabla T_e] - g(T_e - T_l) + aN \tag{S37}$$

$$C_l \frac{dT_l}{dt} = \kappa_l \nabla^2 T_l + g(T_e - T_l) + bN + k_{DS-l}N_{DS} \tag{S38}$$

$$\frac{dN_{BS}}{dt} = -k_{BS-DS}N_{BS} + Q_{exc} \cdot A(t) \tag{S39}$$

$$\frac{dN_{DS}}{dt} = k_{BS-DS}N_{BS} + k_{N-DS}N - k_{DS-l}N_{DS} \tag{S40}$$

where the first three corresponds to the eTTM model correlating the dynamic equilibrium between the energy density of non-thermalized electrons ($N$), effective hot electron temperature ($T_e$), and the lattice temperature ($T_l$), modified to include the rate constant about direct dissipation of non-thermalized electron into excitonic DS ($k_{N-DS}$) and back injection as the heat source to the lattice ($k_{DS-l}$), as inferred from the fluence-dependent mechanistic study (Supplementary Note 9). Other two corresponds to the dynamics of exciton BS and DS, incorporating the counterpart of the plasmon-exciton energy exchange terms present in eTTM equations. All the known parameters consisting of the equation were taken referring to the literature[14], and rate constant $k_{N-DS}$ and $k_{DS-l}$ were modeled empirically, based on the order of magnitude of the timescales reported in previous study.

## Supplementary Note 11. Helicity-resolved mapping of near-field coupling characteristics onto ultrafast dynamics

To comprehensively disentangle the spatiotemporal and polarization-dependent effects—thereby effectively extracting the near-field contributions from the far-field TA signals—we utilized two synthetically tailored variants, $Helicoid_{(A)}$ and $Helicoid_{(B)}$. These structures are designed to exhibit mutually identical steady-state extinction spectra for RCP and LP, respectively (Figure S18a, see Supplementary Note 1). As revealed in Figure 3d and schematically illustrated in Figure S18b, although probing these variants with RCP and LP yields identical macroscopic spectral responses, RCP selectively interrogates the gap region, whereas LP addresses the overall particle morphology.

Surprisingly, when probing TA of $Helicoid_{(A)}$ with RCP, significantly accelerated dynamics appear compared to $Helicoid_{(B)}$ probed with LP, particularly at the LB, despite their identical steady-state far-field spectra (Figure S18c and d). Given the identical macroscopic coupling regime in total particle level, this dynamic divergence must originate from microscopic near-field effects selectively controlled by the incident polarization channel. Specifically, RCP incidence preferentially addresses the chiral gap, allowing us to selectively observe the localized, accelerated relaxation into exciton DS confined within this region (Figures 4g and 4h). This observation is in stark contrast to the dynamics of $Helicoid_{(A)}$ probed with LP, which are highly similar to those of $Helicoid_{(B)}$ probed with LP regardless of their differing steady-state extinction profiles (Figures S18c and e). This perfectly exemplifies the profound advantage of our system: utilizing the helicity of light to selectively address specific sub-diffraction spaces, thereby enabling nanoscale spatiotemporal control over both coupling physics and ultrafast dynamics in plexcitonic systems.

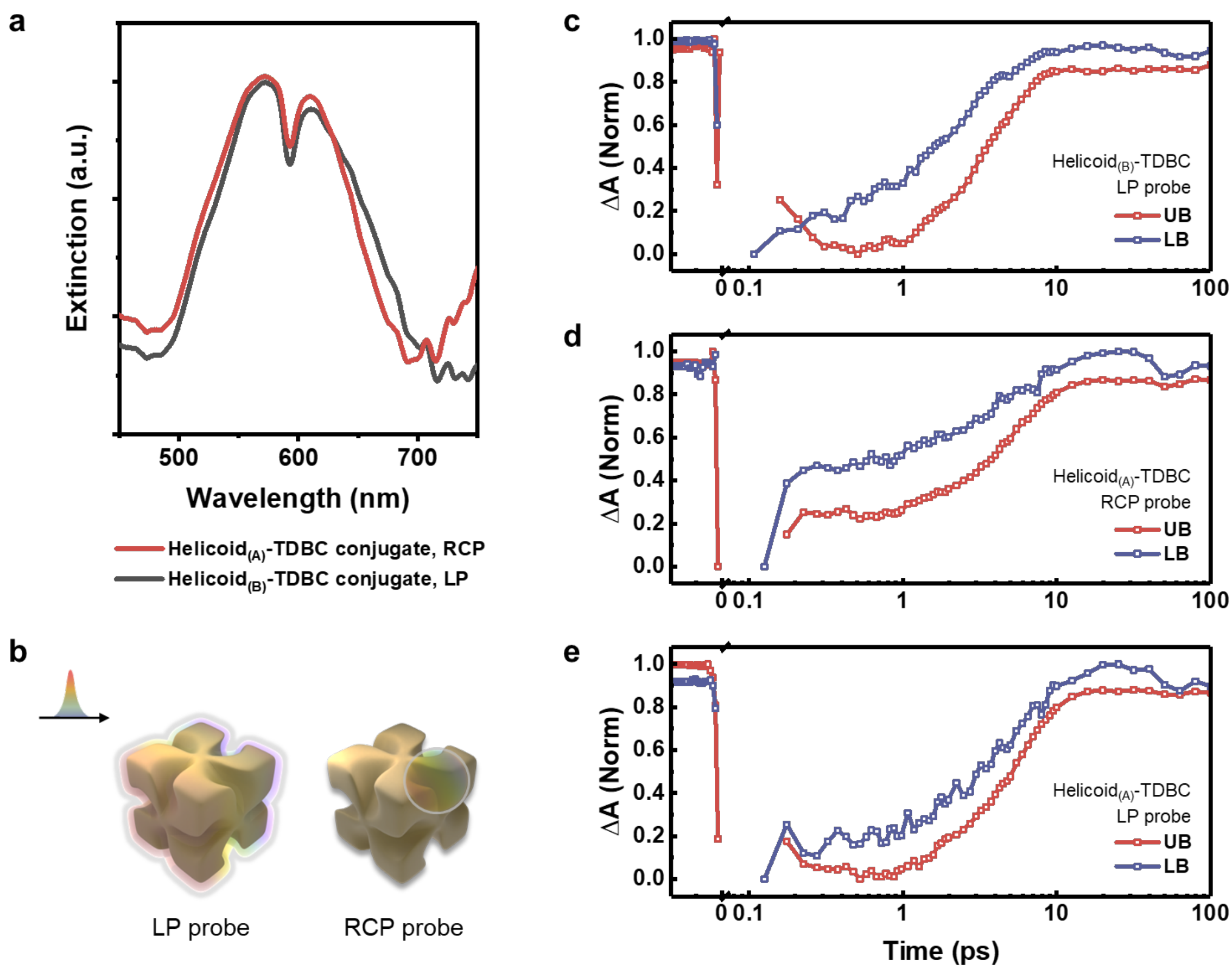

**Figure S18. Helicity-resolved mapping of near-field coupling characteristics onto ultrafast dynamics. a,** Extinction spectra of two helicoid variants (Helicoid$_{(A)}$ and Helicoid$_{(B)}$, also discussed in Figure S1) synthetically designed to have identical spectral response for RCP and LP, respectively. **b,** Schematic illustration of the polarization-resolved selective near-field addressment by the probe pulse. Compared to LP, RCP preferentially focuses within the gap, enabling highly localized access. **c-e,** Ultrafast decay kinetics associated with the UB and LB states of the conjugates. Strikingly, Helicoid$_{(A)}$ probed with RCP demonstrates substantially accelerated dynamics relative to Helicoid$_{(B)}$ probed with LP, even though their steady-state spectra are virtually identical. In contrast, the dynamics of Helicoid$_{(A)}$ probed with LP closely resemble those of Helicoid$_{(B)}$ probed with LP.